\documentstyle[]{mn}
\def\deg{{$^{\circ}$}}

\title{Structures Produced by the Collision of Extragalactic Jets with Dense Clouds}

\author[S.W.\ Higgins, T.J.\ O'Brien\& J.S.\ Dunlop]
{ S.W.\ Higgins$^{1,3}$, T.J.\ O'Brien$^1$ \& J.S.\ Dunlop$^2$\\
$^1$ Astrophysics Research Institute,
Twelve Quays House, Egerton Wharf, Birkenhead L41 1LD, UK\\
$^2$ Institute for Astronomy, University of Edinburgh,
Blackford Hill, Edinburgh, EH9 3HJ, UK\\%[2mm]
$^3$ Current address: Dept. Computing and Mathematics, 
Manchester Metropolitan University,\\
 John Dalton Building, Chester Street, Manchester,  UK\\
e-mail: {\tt S.Higgins@doc.mmu.ac.uk}, {\tt tob@astro.livjm.ac.uk}, {\tt jsd@roe.ac.uk}}

\begin{document}

\maketitle

\begin{abstract}
We have investigated how several parameters can affect the results of a
collision between an extragalactic jet and a dense, intergalactic cloud,
through a series of hydrodynamic simulations. Such collisions are often
suggested to explain the distorted structures of some radio
jets. However, theoretical studies of this mechanism are in conflict
over whether it can actually reproduce the observations.

The parameters are the Mach number, and the relative densities of the
jet and the cloud to the ambient medium. Using a simple prescription we
have produced synthetic radio images for comparison with observations. 
These show that a variety of structures may be produced from simple
jet-cloud collisions. We illustrate this with a few examples, and
examine the details in one case. In most cases we do not see a clear,
sustained deflection. Lighter jets are completely disrupted. The most
powerful jets produce a hotspot at the impact which outshines any jet
emission and erode the cloud too quickly to develop a deflected arm. It
appears that moderate Mach numbers and density contrasts are needed to
produce bends in the radio structure. This explains the apparent
conflict between theoretical studies, as conclusions were based on
different values of these parameters.  Shocks are produced in the
ambient medium that might plausibly reproduce the observed alignment of
the extended emission line regions with the radio axis.
\end{abstract}
\vspace{-3pt}

\section{The effect of environment on extragalactic jets}

\subsection{Complex and distorted structure in extragalactic radio 
sources}

The jets and hotspots of radio galaxies and quasars often show
complex structure, including bends, twists, knots and multiple
hotspots. Such structure is seen over a huge range of sizes, from the
enormous wide-angle tail (WAT) sources (which can be megaparsecs in
size), to the compact steep-spectrum (CSS) sources (which are about
10--15~kpc across). Jets can appear to bend by over 90$^{\circ}$ and
remain collimated for several jet radii (Bridle \& Perley 1984),
despite the expectation that the oblique shock causing the bend
should decelerate the jet. Explanations for these complex structures
include motion through some intra-cluster medium (Leahy 1984),
perturbations due to mergers (van Breugel et al.\ 1986, Sakelliou,
Merrifield \& McHardy 1996), variations in the direction of the jet
at its source (Williams \& Gull 1985, Scheuer 1982), and collision
with dense clouds in the ambient medium (Burns 1986).  Cloud
collisions are particularly applicable in cases where these bends are
very sharp.  Other models would produce more gradual bends.
  
Theoretical studies of jet-cloud collisions appear to be in conflict
over whether they can explain the observations. In some previous
numerical simulations the jet is decelerated and effectively disrupted
by the cloud, and the cloud is subsequently destroyed
(De~Young~1991). In others the jet appears to remain collimated as it
is re-accelerated in a new direction (Norman~1993). Most recently
Raga~\&~Canto~(1996) find from two-dimensional simulations and
analytical studies that the jet bores through the cloud but reaches a
steady configuration. In this paper we describe the results of a study
aimed at resolving this question by investigating the effects of
various hydrodynamical and geometrical parameters. This investigation
uses three-dimensional, adiabatic simulations.  We also follow the
development of the interactions for a longer time than previous
simulations, and estimate the intensity of synchrotron emission from
the jet to enable meaningful comparison with observed radio maps.

Through these studies we hope to be able to determine what types of
complex structure can be explained by jet-cloud collisions. Such studies
could shed light on the alignment between radio and optical axes in high
redshift radio galaxies, the contribution of shocks to the spectra of
extended emission lines regions and the
role of environmental effects in unified schemes
for AGN and their evolution.

\subsection{Examples of distorted structure}

Wide-angle tail (WAT) sources are found in rich clusters and have
distorted, C-shaped structure. Large WATs approaching 1~Mpc across
cannot be bent by the motion of the parent galaxy through the cluster
without assuming unreasonably high speeds through unreasonably dense
intra-cluster gas (Burns 1986). To satisfy the requirements of
momentum balance and to reproduce the sharpness of the bends that are
observed, Burns proposed that the jets may collide with clouds of
higher density in the surrounding medium.  A good example is the
western jet of the WAT 1919+479, which emerges from the core, vanishes
after a short distance, and then reappears at a bright hotspot (Burns
et al.\ 1986, Pinckney et al.\ 1994).  Beyond this hotspot a tail
stretches out for 800~kpc in a direction about 90$^{\circ}$ to that of
the original jet, broadening and fading as it does so. The bend here
is very sharp.  There is one other hotspot not far from the beginning
of the tail. Rotation measure and depolarisation are described as
`patchy', and vary significantly over the tail. X-ray observations
show large scale asymmetry in the intra-cluster medium.  The
simulations of Loken et al.\ (1995) show that the necessary gas
velocities can arise in cluster mergers, as can shocks that will bend
the jet. However, these simulations still do not explain the sharpness
of the bend or the persistence of the halo.

Barthel et al.\ (1988) present a large sample of quasars with powers greater
than the Fanaroff-Riley division but distorted structures. Twenty out of a
sample of eighty high-redshift quasars showed bending greater than 20\deg,
instead of the classical double structure that would be expected. 

Compact Steep Spectrum (CSS) sources make up 10--15\% of AGN. More
than 15\% have axes between the lobes and the core that 
are misaligned by more than 20$^{\circ}$ (Saikia et al. 1995). This 
distortion seems to be associated with an asymmetrical ambient medium.

In quasars the most prominent jets are in complex or one-sided structures,
and smaller sources are the more powerful (Muxlow \& Garrington 1991).
Stocke, Burns and Christiansen (1985) present observations of `dogleg'
quasars showing strong changes in direction within the lobes. 
 There is no evidence that
these sources are found any more often in rich or poor cluster environments,
a conclusion supported by recent work (Rector, Stocke \& Ellingson
1995). Only a few dense clouds are needed in each case to explain the
proportion of bent jets, so it would appear that it is the inhomogeneity of
the environment, not its overall density, that causes the bends.

\subsection{Intergalactic clouds in the neighbourhood of
extragalactic jets}

Evidence for inhomogeneity in the medium surrounding AGN comes from
measurements of depolarization and line emission, and correlations
between these properties/features and the radio structure.

Maps of rotation measure and depolarization of a sample of radio sources
show significant inhomogeneity on a range of scales, some as low as 5kpc,
some larger than 50kpc, out to distances of about 100kpc from the nuclei
(Pedelty et al.\ 1989). Such effects could be caused by density
inhomogeneities in the surrounding material.

The infrared, optical and ultraviolet structures of many high-redshift
radio galaxies are closely aligned with the radio structures (McCarthy
 1993, Chambers, Miley \& van Breugel 1987). There are also
several correlations between line emission and other aspects of
distorted jets, for example: brighter line emission occurs on the side
of the nearer radio-lobe (McCarthy, van Breugel \& Kapahi 1991); the
level of blue light and the strength of the alignment effect are
correlated with a mix of radio power and spectral index (Dunlop \&
Peacock 1993). The lobes of these radio sources are often asymmetrical
with the lobe nearer the nucleus tending to be more depolarized than
the more distant lobe (Liu \& Pooley 1991), suggesting that the material
responsible for the depolarization may also present more resistance to the
jet.

Regions of optical line emission are associated with knots and bends in the
jets (Wilson 1993). For example, the radio galaxy 4C 29.30 has a region of
line emission close to a bright knot just before a bend (van Breugel et al.\
1986).  The radio galaxy PKS2250-41 shows extended emission aligned
with the radio axes, including a large arc-shaped region of line emission
surrounding the radio lobe (Tadhunter et al.\ 1994).
Extended emission line regions have a ``clear spatial association'' with
regions of depolarized radio emission (Baum \& Heckman 1989).  

In many sources (especially at higher redshifts) the spectra from this
extended emission line region can only be explained if a significant
shock component is included (Clark \& Tadhunter 1996, and references
therein). Low polarization of the ultra-violet emission shows that
there is not enough scattered AGN emission to account by itself for
the total flux (Tadhunter 1996).  The cloud collision model is now
commonly invoked to explain the properties of extended emission line
regions in AGN, for example the Seyfert galaxy NGC 1068 (Axon 1996);
radio galaxies 3C 254 (Crawford 1996) and 3C368 (Stockton, Ridgway \&
Kellog 1996); and Cen A (Sutherland, Bicknell \& Dopita 1993), as well as
the sources mentioned in the previous paragraph.

 The most luminous radio sources are known from observations to reside in
regions of high galaxy density (Yates, Miller \& Peacock 1989, Hill \& Lilly
1990). X-ray observations of powerful radio galaxies and quasars show that many
lie in the centres of rich clusters with dense, rapidly cooling IGM in which
cold clouds can condense (Fabian 1993). Cowie et al. (1983) observed
filamentary line emission in cooling flows that could indicate the presence of
such clouds. Observations suggest that these clouds have temperatures of about
10$^4$ K, densities of about 100 cm$^{-3}$ and sizes of 3--15kpc (Baum 1992).
Some contribution from radio jets seems necessary to re-energize the filaments.

\section{Simulating collisions of jets and clouds}

\subsection{Previous simulations}

Analytical studies show that sharp bends of 90$^{\circ}$ are possible
if the jet is thin or only moderately supersonic (Icke 1991). More
recently Raga \& Canto (1996) have published analytical calculations
and two-dimensional simulations showing bending by clouds. They
conclude that slower jets will be bent more, and clouds will be eroded
as jets bore through them.

The first investigation of the effect of off-axis jet-cloud collisions
was by De Young (1991) using the `beam scheme' (Sanders \& Prendergast
1974), to test the proposal of Burns (1986) that bending in large WATs
is the result of collisions with clouds. De Young monitored the jet
flow using test particles, and observed that the jet was considerably
decelerated by the impact with the cloud.  The cloud was destroyed
within a few million years, and the jet returned to its original
direction.  He concluded that the interaction does not last long
enough to produce anything like a tail.

However, WAT tails have a wide opening angle and show no strong evidence of
supersonic speeds (such as terminal hotspots). It would appear the jet is
disrupted at the impact point anyway. The only question is whether the
interaction can be maintained long enough to produce tails of the observed
length.

A similar interaction was investigated at higher resolution by Balsara \&
Norman using their RIEMANN code (Norman 1993). They argued from plots of the
velocity field that a De Laval nozzle was formed which re-accelerated the
jet in a new direction after impact.

We aimed to resolve this conflict through the work described in this
paper, and to more thoroughly investigate the effect of
different parameters both on the development of the interaction over
time, and the structures produced, and by estimating the appearance of
the source at radio wavelengths for direct comparison with
observations.

\subsection {Numerical techniques}

The simulations we present in this paper were performed using a
hydrodynamic code based on the Godunov method of Falle (1991) in
three-dimensional Cartesian coordinates. This technique solves the inviscid
Euler equations to second-order accuracy in space and time, with an
adiabatic equation of state. In addition to calculating the usual dynamical
variables, we also calculate a parameter representing the fraction of
density within each cell which was originally jet material. This allows us
to follow the evolution of the jet separately from the ambient medium and
to calculate synthetic radio maps as described in section \ref{synchro}.

\begin{figure}
\vspace {18.4truecm}
\includegraphics{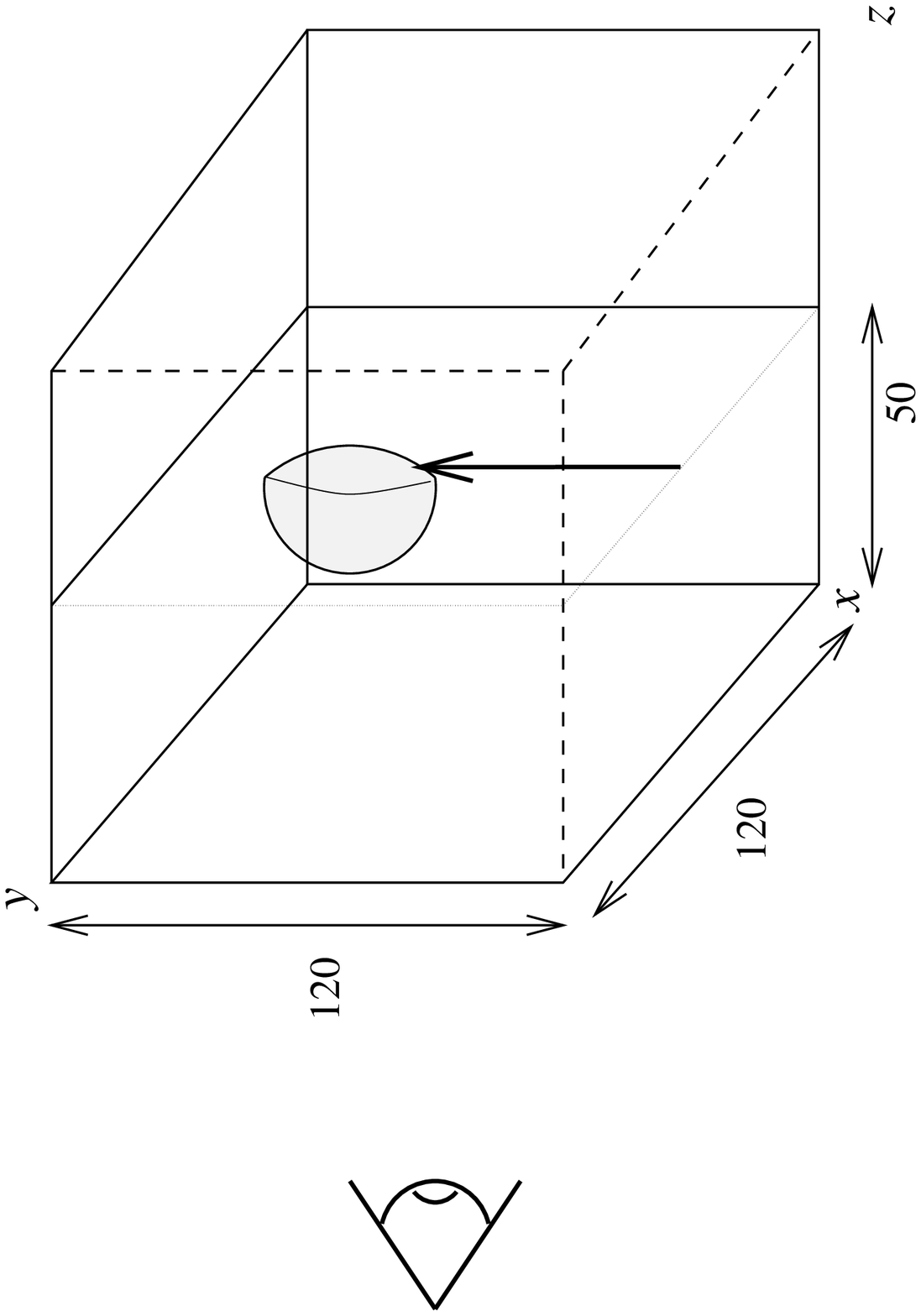}
\vspace {0.5truecm}

\caption{%The geometry of the computational grid. This shows the symmetry plane
% containing the jet axis and bisecting the cloud. It also shows the viewpoint
% used for the density slices in figures \ref{hirestest} to \ref{jet7}. 
}
\label{symfig}
\end{figure}

Since we can draw a plane of symmetry bisecting the cloud and containing
the jet we have only calculated one half of the region around the
interaction (figure \ref{symfig}).  The boundaries of this domain are
treated as free flow except for the symmetry plane and the region where the
jet enters the grid. A free flow condition assumes that values on the outer
surfaces of each cell are exactly equal to those on the inner surfaces. The
symmetry condition is that velocities normal to the surface are reflected.
The jet is produced by using an appropriate boundary condition representing
incoming material in the region where the jet enters.  Note we have made no
assumptions about the position of the central engine with respect to the
grid.

The simulations can be rescaled so that they represent structures on parsec
or kiloparsec scales, as long as the gas behaves adiabatically. The
simplest rescaling is to change sizes and times in proportion, preserving
all other variables. For example, if the cell side is 1 kpc, then 1 time
unit = 2$\times 10^6$years.  We use this scaling as a reference in
discussing the simulations below.  We can apply these simulations to other
cases with different pressures and temperatures. For example, in the case
of the fastest jets, the temperature may be a hundred times higher, or for
the slow jets a hundred times lower, and we still have velocities in the
range accepted for extragalactic jets.

\subsection{Testing the hydrodynamic code}

We used the code to calculate two one-dimensional test problems. The
first was Sod's shock tube problem (Sod 1978). The code was used to
compute plane shocks moving along each of the three axes of the grid
(in both forward and reverse directions, as well as at various
resolutions). It produced results in good agreement with the analytical
solution.

\begin{figure}
\vspace{20truecm}
\includegraphics{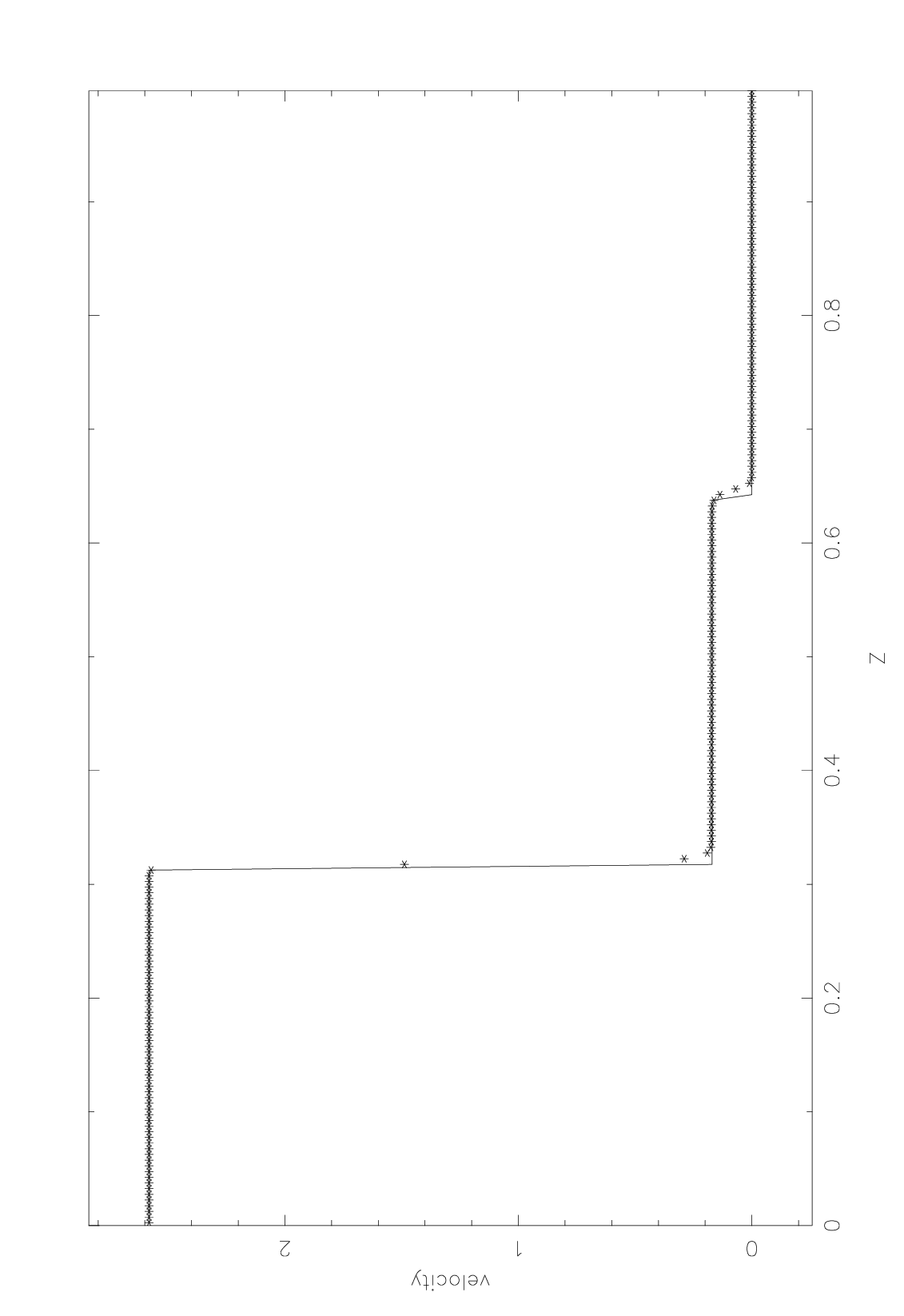} 
\includegraphics{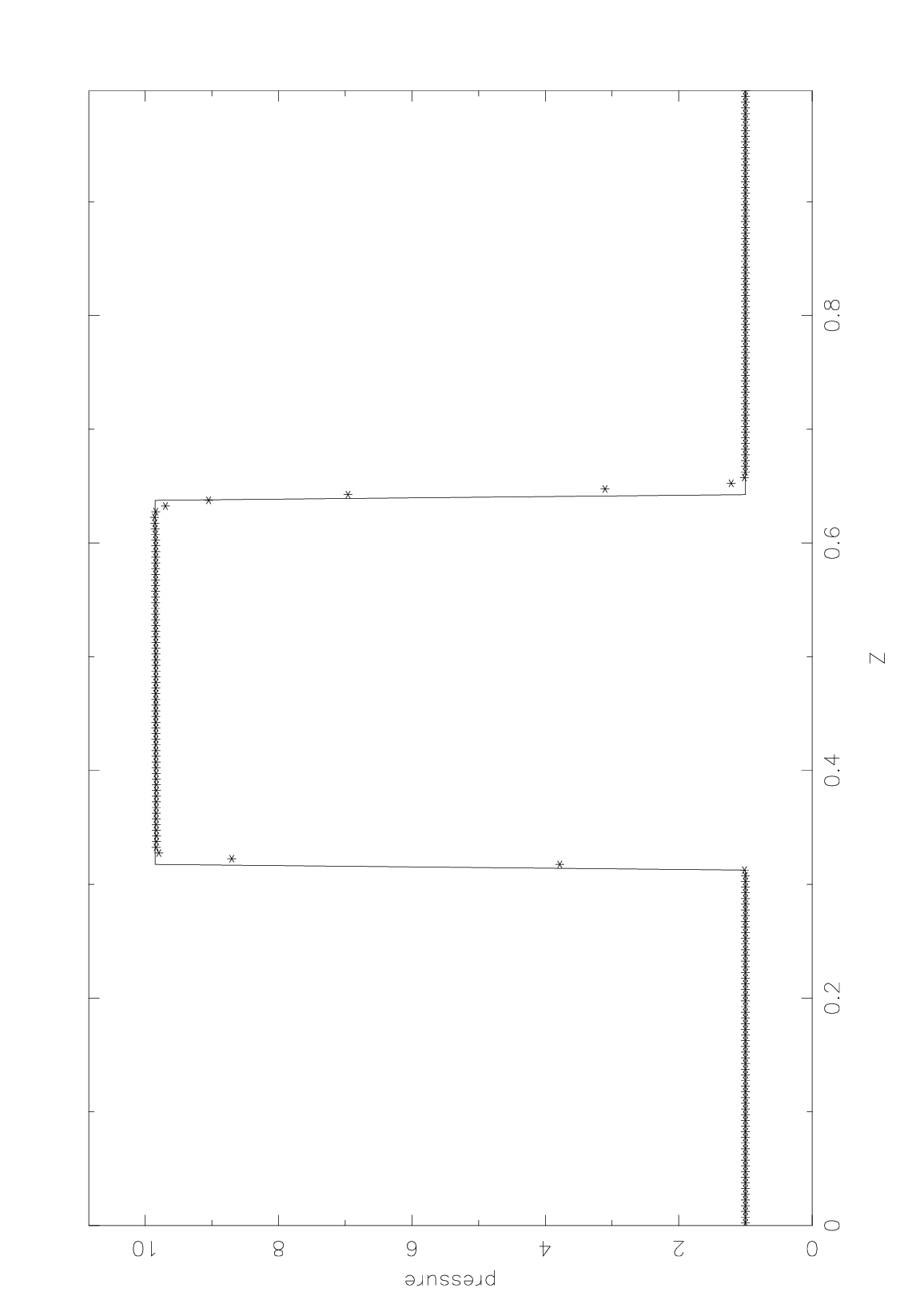} 
\includegraphics{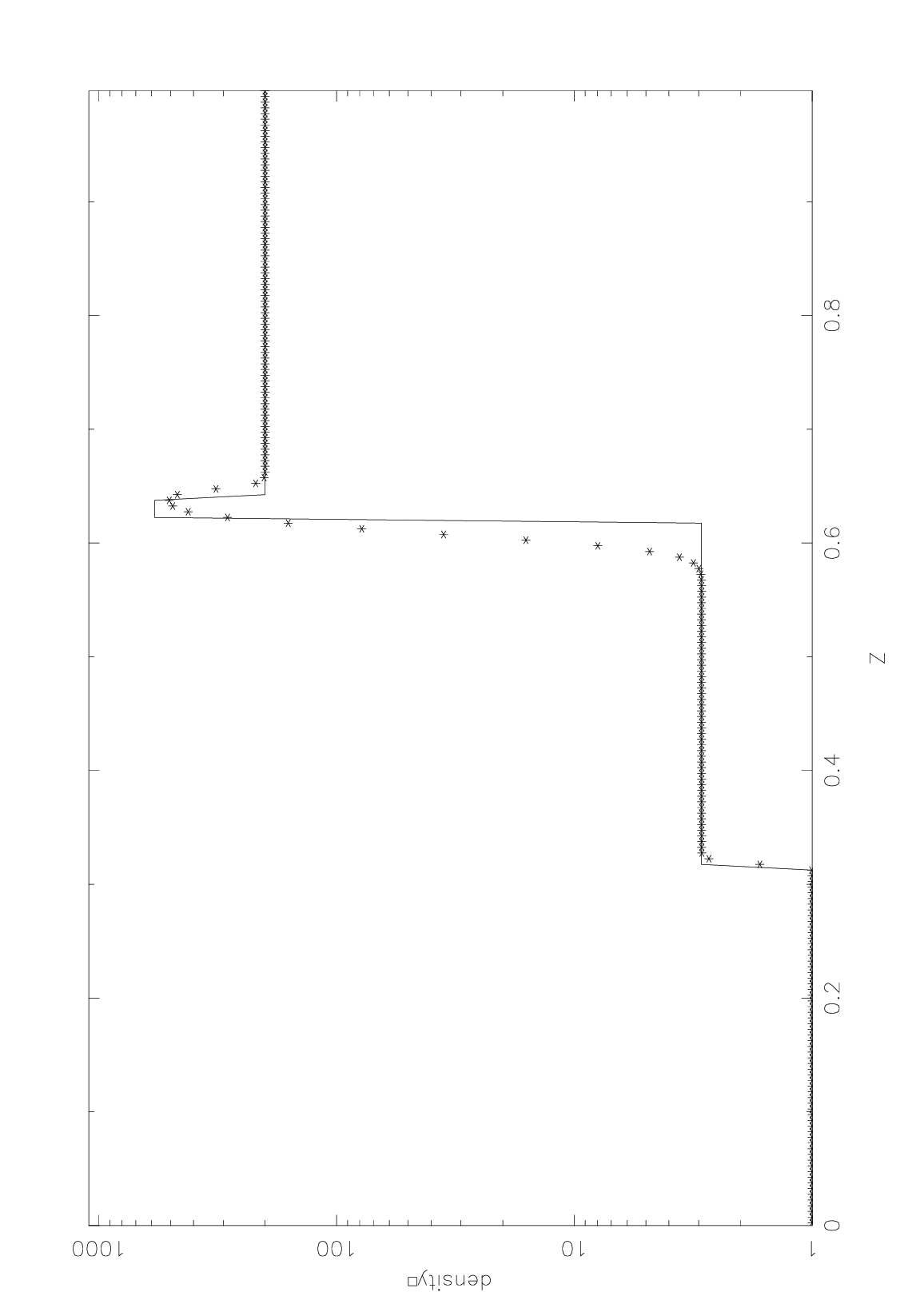} 
\vspace{2cm}
\caption{}
\label{test1}
\end{figure}

The second test problem was the collision of a one-dimensional shock with a
density discontinuity (Nittman, Falle \& Gaskell 1982). This problem is
clearly appropriate to our investigation as a useful indication of the
fidelity of the code in this case. Once again this was run with shocks and
discontinuities normal to each of the three axes, moving both forwards and in
reverse directions and at various resolutions. Figure \ref{test1} shows an
example of the results (full details are given in the caption to the figure).
The positions and values of the two shocks were accurately reproduced.

As a final test we re-calculated a portion of one of the simulations
(simulation 3 -- see section \ref{sumres}) at double the initial
resolution within a volume one-eighth of the size and compared the
results. When the high-resolution results were smoothed to the lower
resolution they showed no significant differences. Nor did there
appear to be any effects caused by allowing the simulated flow to
leave this grid compared to the same region of the lower resolution
simulation.  In figure \ref{hirestest} we show density slices at the
same time for the simulations at both resolutions.  We are satisfied
from this that the results of our simulations are not significantly
affected by the resolution or boundary effects.

\subsection{Producing synthetic radio maps}

\label{synchro}

Previous studies of jet-cloud collisions relied on the interpretation of
flow patterns produced by the simulations to reach their conclusions. To
allow a more direct comparison we have developed a simple approximation
for the intensity of synchrotron emission in terms of the hydrodynamic
variables. We use this to produce estimates of the surface brightness
distribution of radio emission. These synthetic radio maps can be
compared with observations. In this section we describe our
prescription, and the assumptions it is based on. We then consider how
changing these assumptions might affect the results, and present some
plots for alternative prescriptions.

The intensity $j$ of synchrotron emission at frequency $\nu$ is given by
$j \propto K B^{1+ \alpha}\nu^{-\alpha}$, where $B$ is the magnetic
field strength, $\alpha$ is the spectral index and $K$ is related to
the number density $N(\gamma)$ of relativistic electrons with Lorentz
factors in the range ($\gamma$,$\gamma+d\gamma$) via $N(\gamma)d\gamma =
K \gamma^{-(2\alpha+1)}d\gamma$. In order to calculate $j$ we need to
express the magnetic field and the coefficent $K$ in terms of the
results of our hydrodynamic simulations.

To determine the magnetic field we  assume that the field is
insignificant outside the jet (see, for instance, Smith et al. 1985) and the
magnetic flux is frozen into the jet material. By conservation of magnetic 
flux we would expect the field strength to be related to the jet density 
via $B \propto \rho_{jet}^{2/3}$.
In practise  a complicated flow can amplify an initially disordered
field. For the purposes of this work we have neglected this amplification.

To determine the coefficient $K$ we assume that the energy density of
relativistic particles is a fixed fraction of the internal energy
density of the gas with the only changes in the distribution being due
to adiabatic expansion. This is equivalent to assuming that the
relativistic electrons form a supra-thermal tail throughout the gas,
and the efficiency of the acceleration process is similar
everywhere. Following Wilson and Scheuer (1983) we can relate $K$ to
the gas pressure $p$ via $ K \propto p^{(\alpha/2+3/4)}$. The
adiabatic index varies between 5/3 for non-relativistic gas to 4/3 for
relativistic particles which corresponds to an uncertainty of about
0.3 in our choice of $\alpha$.

We substitute for $K$ and $B$ from the formulae above, and therefore
obtain

\begin{equation}
j\nu^\alpha\,\propto\,p^{\alpha/2+3/4}\rho_{jet}^{2\alpha/3+2/3}\
\end{equation}

With $\alpha$=0.5 we have $j\nu^\alpha \propto p \rho_{jet}$, whereas
with $\alpha = 1.0$ we find $j\nu^\alpha \propto
p^{5/4}\rho_{jet}^{4/3}$.

As an alternative to assuming that magnetic flux is frozen into the jet
and that it is zero elsewhere we could assume that the magnetic energy
density is in equipartition with the internal energy density of the gas
(and, because of our other assumption, with the relativistic particle
energy density). This leads to $B\propto p^{1/2}$ and therefore

\begin{equation}
j\nu^\alpha\,\propto\,p^{\alpha+5/4}\
\end{equation}

It is common to use $\alpha$=0.75 so that $j\nu^\alpha \propto p^2$.

Further, if one still believed that the magnetic field was negligible
outside the jet then we could use this expression with one's favourite
$\alpha$ wherever $\rho_{jet}\neq 0$ and set the intensity equal to zero
where $\rho_{jet}=0$. 

We can calculate the synchrotron intensity for each computational cell and,
assuming that the emission is optically thin, integrate along lines of
sight to produce a synthetic radio map at any epoch. In figure
\ref{rad_comp} (a -- d) we show four alternative maps from the same fluid
conditions based on the alternatives described above (see caption). Using
the square of total pressure, the hotspot is large with a shape like a
tadpole, but neither the jet nor the deflected tail are detectable. 
However, the evidence for higher magnetic fields in the jet is strong.
The other alternatives produce
structures that are similar to each other, so that the general comparisons
we would like to make are not seriously effected.

\section{The effect of jet and cloud properties on the results of the 
interaction}

\subsection{Models and coverage of parameter space}

The parameter space of jet cloud collisions is multi-dimensional. In an
attempt to explore the range of behaviour within this space we have
chosen to vary the values of three of the most significant parameters
over ranges applicable to extragalactic jets: the jet Mach number, its
density contrast with the ambient medium, and the contrast between cloud
and ambient density. Using one of these cases, we have also investigated
the effect of impact angle and relative size of cloud and jet on one of
the more interesting cases. Details are given in table \ref{pstbl}.

We have assumed conditions in the ambient medium consistent with
observations, that is a temperature of 5$\times 10^7$K and a particle
number density of 0.01 cm$^{-3}$. These values are used to normalize
the quantities in the computation so that model values for the ambient
density and pressure in the code are set to 1.0. The jet and cloud are
both taken to be in pressure balance with the ambient medium. The
computational domain is divided into a grid of 50 $\times$ 120
$\times$ 120 cells.

\subsection{Summary of results}

\label{sumres}

By simply changing a few parameters we have produced a range of different
structures. We discuss these in this section. The common features in all
the simulations are: the jet is disrupted and decelerated to some extent on
collision; the cloud is eroded by the jet; high pressures and densities are
produced at the point of impact, giving rise to bright hotspots in the
radio emission. All these structures vary in time, and many might be
associated with features observed in real radio sources at different epochs
in the simulation (Higgins, O'Brien, \& Dunlop, 1995).

We illustrate this discussion by presenting some examples. In each
case we show total density beside a synthetic radio map at two epochs.
Density plots show a slice through the symmetry plane (see figure
\ref{symfig}), rendered in a logarithmic greyscale. The synthetic
radio maps are produced by estimating the intensity of synchrotron
emission in each cell (using the formula $j\nu^\alpha \propto p
\rho_{jet}$, as described in section \ref{synchro}). This is then
summed along the line-of-sight perpendicular to the symmetry
plane. The resulting maps are shown on a logarithmic greyscale with a
range of about one order of magnitude. They have very different peak
values, so the hostpots are not directly comparable, but the dynamic
range between the peaks and faintest features are.  The faster jets
have left the grid in all cases before the slower jets have developed
any interesting features, so we show each jet shortly after impact,
and then at a later stage of the interaction.  For the fast jet this
is $t=$4 units of computational time after it enters the grid (16
million years on the scaling specified above) and $t=$28 (56 million
years). The slower jets are also shown at $t=$28, which is around the
time of impact, and then at $t=$76 (112 million years).

The light, slow jet (simulation 1, figure \ref{jet1}) is scattered into
a very broad area in all directions perpendicular to the jet. A
mushroom or umbrella shaped structure is formed in the radio
emission. The hotspot is no brighter than it was in the ambient medium
before impact, and is recessed a little within the lobe. It is not
obvious from the radio structure that this is a cloud collision. 
 When the same jet encounters a denser cloud (simulation 2) the interaction
lasts longer, since the jet erodes the cloud much more slowly. The
denser cloud makes no difference to the hotspot behaviour, which is
not significantly brighter on impact with the cloud. This feature
almost certainly depends on the relative sizes of the jet and cloud: a
much larger cloud would not allow deflection in both directions.

The fast, light jet (simulations 3 and 4, figure \ref{jet4}) produces
a weak secondary hotspot at the head of the deflected jet in the radio
map. This is about a tenth as bright as the spot at the impact
point. It becomes disconnected (on the dynamic range of our radio
plots) and disappears within about fifty thousand years, and the jet
rapidly erodes the cloud (whatever its density).  As the jet breaks
through the cloud there are two hotspots within a boot-shaped lobe.
Although the cloud impact is the cause of the bending, the deflection
and secondary hotspot is actually produced as the jet bends inside the
distorted cocoon that has been formed during the interaction.

 After a few hundred thousand years it has broken
through the cloud and shot off the grid. This faster jet can be
deflected, and produces much more visible results, but only for a
short time.  In this case the hotspot shows a small increase in
brightness (about 10~--~20\%) on impact, and remains about the same
throughout the interaction.
The density of the cloud does not make significant
difference to the effect on the jet 
in either of these cases.

The slow, heavy jet (simulations 5 and 6, figure \ref{jet7}) shows a
clear deflection by 90$^{\circ}$. The jet forms a bowl-shaped cavity
in the cloud which deflects the jet with a relatively large radius of
curvature. We do not see any secondary hotspot in the radio map. The
primary hotspot brightens steadily after impact, up to a factor
between two and three. The jet erodes the cloud slowly. On the scale
of a WAT source it takes about $3 \times 10^8$ years to erode the less
dense cloud.  There is a hotspot down the jet before the impact point.
When the cloud is denser the interaction does of course last
longer. The deflection angle is nearer to 70$^{\circ}$. After about
6$\times 10^8$ years the deflected arm is about four jet radii long,
and about half as wide again as the incoming jet. The jet has worked
about half way through the cloud. The shape of the deflecting face
formed by the interaction is now deeper, so disrupting the jet
more. However the interaction can certainly continue for some time
yet.

The fast, heavy jet (simulations 7 and 8) pushes deep into the cloud,
however dense it may be. The impact hotspot is orders of magnitude
brighter than any other features.  The jet breaks through the cloud
within about 50 million years. The radio map shows an extremely bright
deflection spot and weaker secondary (by 1000 times) in the deflected
material. The primary hotspot brightens by a factor of ten on impact
in this powerful jet. Because this interaction is short-lived, and the
hotspot far outshines any other emission, it would be difficult to
detect any deflection in a source with this powerful a jet.

\subsection{The Effect of Impact Angle}

We used the same fluid parameters as in simulation 4, altering only
the angle at which the jet encountered the cloud. With an impact angle
of 45$^{\circ}$ this produces a distinct secondary at the head about
one jet radius away.  However, it is shorter lived ($\sim$ 40
million years) and the angle of deflection decreases quickly as the
jet erodes the cloud.  With an impact angle of 30$^{\circ}$ the
deflection is shallower ($\sim 70^{\circ}$) and shorter
lived. Shallower impact angles can produce secondary hotspots since
less momentum is lost by the jet, but the interaction is short lived
on the whole. An impact angle of 45$^{\circ}$ is probably the lower
limit for a deflection of 90$^{\circ}$.

\subsection{An Example of a Jet-Cloud Interaction}

Our sixth simulation ($M$=2, $\eta$=0.2, figure \ref{jet7}) shows all
the major features of WAT jets: bent sharply through 90$^{\circ}$ at
bright hotspots and flaring out into long, wide tails (O'Donoghue,
Eilek \& Owen 1990). Due to the higher cloud density the interaction
is long-lived. It lasts for something of the order of 10$^8$ years,
which allows the formation of a tail not only large enough but of an
age consistent with estimates of travel time to the end of the tails
from synchrotron spectral-aging (O'Donoghue, Owen \& Eilek 1993).

We chose this simulation to examine the dynamics of
the interaction in more detail. In particular we  look at the
distribution of material and its velocity in three dimensions, along with
the estimated
radio emission, and how it varies with viewing angle. We also give some
indications of the likely location and shape of line emission.

Figure \ref{jet7_1} shows velocity vectors plotted in three dimensions, with
the initial jet direction emerging from the plane of the diagram. This is at a
later time than shown in figure \ref{jet7} (3$\times 10^8$years). These show
how the material is deflected after impact into a fan of opening angle between
60$^{\circ}$~--~80$^{\circ}$. None of this material is moving slower than 20\%
of the speed of the incoming jet. A spine of slightly faster material can be
discerned along the middle of this fan, representing the remnant of the
jet. The wings are material deflected at higher speed on impact; as the
interaction progresses the impact cavity deepens and material deflected in this
direction at later times has slightly lower speeds that fall below the
threshold of this plot. Clearly on deflection the jet is not only decelerated
but almost all collimation is lost. However, this fan is very flat, allowing
the appearance of a deflected tail when viewed from the side.

The next plots (figure \ref{rad}) show the expected radio emission, viewed from
two orientations.  Since the fan is fairly thin this gives a
reasonable impression of a bent jet from the side (a), with the tail
showing many similarities to WATs, including filamentary structure.
 This seems to be due to vortex action in the tail. In (b) we
see how this radio structure would appear were it not close to the
plane of the sky. The emission from the deflected fan now surrounds
the whole jet. This emission should not be as bright as the previous
case since that had longer path lengths through the emitting
material. Thus it should be more difficult to identify the
counterparts of these objects at different orientations.

\subsection{Line emission}

The compression and distortion of the cloud would be likely to result in
enhanced line emission, due to photoionization by continuum from a hidden
quasar (if not limited by the flux from the nucleus) or shock heating. The
simulations show elongated density enhancements beside the jet, especially
the transmitted shock driven into the cloud. As a crude indicator of the
expected position and distribution of line emission, we have plotted the
distribution of density squared (emission measure) integrated through the
grid in figure \ref{em}. This is overlayed in the figure with contours of
radio emission. Note the bright, elongated region inside the cloud, forming a
cap around the impact point. Compact regions of line emission are seen in HST
images (for example 4C41.17, van Breugel 1996) which we interpret in this
model as shock fronts moving away from the jet, providing density
enhancements for scattering of or photo-ionization by the AGN continuum, or
shock-ionization. This gives a natural explanation for the radio-optical
alignment effect.

\subsection{The Effect of Several clouds}

In realistic situations there are likely to be many clouds, so we have
simulated the passage of a jet through a medium containing an ensemble of
clouds. Figure \ref{multi} shows a three-dimensional rendering of total density and
a synthetic radio map at the same epoch. As the jet progresses through the
grid it collides with clouds, producing prominent hotspots in the radio
emission. These spots persist as the jet moves past the cloud and encounters
further obstructions. They fade as the clouds are eroded by the passage of
the jet. Meanwhile new hotspots form at new encounters, and deflected jet
material percolates through the ambient medium, producing filamentary and
foamy shock structures. The result is a jet that is made visible by a series
of irregular knots, with a crooked ridgeline, filamentary diffuse bridges
and lobes and multiple hotspots at the head of the jet.
It clearly remains collimated and produces
a bow-shock at its head.

\section{Conditions for the deflection of jets}

Synchrotron emission may not trace the fluid flow in an extragalactic
jet in a simple way. Bends in the radio structure may not represent
bends in the flow.  Our synthetic radio maps allow us to make more
general comparisons between our simulations and observations. De Young
and Balsara and Norman based their conclusions solely on the flow
patterns. Our results show how a few changes in parameter values can
alter the results of a collision. Under our assumptions, it is
possible to produce structures reminiscent of those seen in a variety
of radio sources. This also explains the apparent conflict between
previous simulations. The simulations performed by De Young involved a
fast, heavy jet (about Mach 25, with a density contrast of $\eta$=1)
so it is not surprising that it ripped up the clouds. In contrast,
Balsara and Norman's jet was of a moderate speed and lightness (Mach 4
and a density contrast of $\eta$=0.2). As we have shown it is easier
to produce deflection under these conditions. The conclusions clearly
depend on the choice of parameter values.

The impact causes a large increase in radio luminosity of fast jets,
which can still be seen after tens of millions of years. Samples of
ultra-luminous radio sources may contain a large proportion of sources
in which jets have been in collision with clouds within this time in
their past, even if the probability of a collision is fairly small for
any source. The technique of ultra-steep-spectrum selection used to
locate luminous sources at high redshift (Rottgering 1992) could make
this bias even stronger.

The range of structures we have produced suggest that deflection may
be easier to produce or detect in lower power jets propagating in the plane
of the sky. We attempt to model sources in more detail in future
papers, but note that in the case of WATs these do not seem very
demanding requirements: the sizes are already very large, so we would
not expect any other orientation. When such sources are seen at some
other orientation the lobe emission may be too diffuse to
detect. However, our simulations use spherical clouds, which are
clearly an idealised case.  Other shapes or clouds with some density
variation may produce a better degree of collimation in the deflected
jet. This might allow deflected structures to be observed from a wider
range of angles. Larger clouds may sustain the deflection of high
power jets long enough to produce a deflected arm.

Steffen et al.\ 1997 have studied the interaction of jets with clouds
in the narrow-line region of Seyfert galaxies through two-dimensional,
non-adiabatic simulations. These suggest that only clouds above a
critical density will radiate after being shocked by the impact of the
jet. Thus the absence of detectable emission associated with bends may
be due to a low cloud density.

We have also tentatively associated our other simulations with bent
structure. We discuss this in detail elsewhere. In particular we find
that different structures may be produced by a single set of
parameters as the interaction progresses (Higgins, O'Brien \& Dunlop
1995, Dunlop 1995). Other sources can also be modelled by these
simulations (with rescaling when appropriate). Our studies show that
shallower impact angles can produce secondary hotspots since less
momentum is lost by the jet, although these are short-lived. Thus
compact sources seem to be best modelled with more oblique impacts
and/or several clouds (see figure \ref{jet4}).

\subsection{Further Development}

In conclusion, given the right sets of conditions collisions between
jets and clouds can reproduce some of the distorted structure seen in
observations, and is suggestive of alignments between radio and
optical axes. A more detailed understanding of the physics involved
could allow us to infer the properties of jets and their environment.

The radio maps presented here assume a fixed spectral index throughout
the source. A more sophisticated prescription in which this could vary
would allow us to investigate the variation of the appearance of a
source with frequency. It should also be possible to include
calculation of a few common lines, and their spatial distribution.

These calculations are non-relativistic. Relativistic simulations of
jets show no gross differences to the structures seen in
non-relativistic simulations. The most significant difference is in
the effect of Doppler beaming, which should not apply in our case
since the lobes are not beamed.  Recent calculations suggest that
relativistic jets may lose less kinetic energy through entrainment
of ambient material (Bowman, Komissarov \& Leahy 1996).  If this was the
case in our model it might allow better collimation after deflection.
We intend to  make relativistic calculations to explore
this possibility.

\newpage

\section*{Acknowledgments}

SWH acknowledges the PPARC for receipt of a studentship, and his wife,
Leah, and parents for additional financial subsidies. Computing was
performed using the Liverpool John Moores University Starlink node. We
would like to thank Dr Sam Falle for useful suggestions, and Dr Huw
Lloyd and Dr John Porter for valuable discussions and the referee for pointing
out important considerations.

\newpage
%\noindent
\section*{References}

\parindent=0pt

Axon, D., 1996, in Clark, N.E., ed., Jet-cloud Interactions in 
Extragalactic Nuclei, published on the World Wide Web, \vspace{-18pt}
\begin{verbatim}
http://www.shef.ac.uk/~phys/research/astro/conf/
index.html
\end{verbatim}
\vspace{-8pt}

Barthel, P.D., Miley, G.K., Schilizzi, R.T., Lonsdale, C.J., 1988, 
A\&AS, 73, 515

Baum, S.A., 1992, in Fabian, A.C., ed., NATO ASI Ser., vol. 366, 
Clusters and Superclusters of Galaxies, Kluwer, Dordrecht, p. 171

Baum, S.A., Heckman, T.M., 1989, ApJ, 336, 681

Bridle, A.H.,  Perley, R.A., 1984, ARA\&A, 22, 319

Bowman, M., Leahy, J.P., Komissarov, S.S., 1996, MNRAS, 279, 899

Burns, J.O., 1986, Can.J.P.,    64,    363

Burns, J.O., O'Dea, C.P., Gregory, S.A., Balonek, T.J.,
1986, ApJ, 307, 73

Chambers, K.C., Miley, G.K., van Breugel, W.J.M., 1990, ApJ, 363, 21

Chambers, K.C., Miley, G.K., van Breugel, W.J.M., 1987, 
Nat, 329, 624

Cowie, L.L., Hu, E.M., Jenkins, E.B., York, D.G.,
 1983, ApJ, 272, 29

Clark, N.E., Tadhunter, C.N., 1996, in Carilli, C.L., Harris, D.E.,
eds., Cygnus A~--~Study of a Radio Galaxy, CUP, Cambridge, p. 15

Crawford, C.S., Vanderriest, C., 1996, MNRAS, 285, 580
%in press, Preprint: CAP-9610003

De Young, D.S.,  1991, ApJ,    371, 69

Dunlop, J.S., 1995, in Hippelein, H., Meisenheimer, K., eds.,
Galaxies in the Young Universe, Lecture Notes in Physics vol. 463, 
Springer-Verlag, Berlin

Dunlop, J.S., Peacock, J.A.,  1993, MNRAS,    263,     936

Fabian, A.C., 1994, ARA\&A, 32, 277

Falle, S.A.E.G.,  1991, MNRAS,    250,     581

Higgins, S.W., O'Brien, T.J., Dunlop, J.S., 1996, in 
Ekers, R., Fanti, C., Padrielli, L., 
Extragalactic Radio Sources,  IAU Symp. 175, Kluwer, Dordrecht, 467

Higgins, S.W., O'Brien, T.J., Dunlop, J.S., 1995, in Millar, T.J., 
Raga, A., eds, Shocks in Astrophysics,  Kluwer, Dordrecht, p. 311

Hill, G.J., Lilly, S.J., 1990 ApJ, 367, 1

Icke, V., 1991, in  Hughes, P.A., ed,
Beams and Jets in Astrophysics,  CUP, Cambridge, p. 232

Koide, S., Sakai, J-I., Nishikawa, K-I., Mutel, R. L., 1996, ApJ, 464, 724

Lacy, M., Rawlings, S., 1994, MNRAS, 270, 431

Liu, R., Pooley, G., 1991, MNRAS,  249, 343

Leahy, J.P., 1984, MNRAS, 208, 323

Leahy, J.P., Muxlow, T.W.B., Stephens, P.W. 1989, MNRAS, 239, 401

Loken, C., Roettiger, K., Burns, J.O. Norman, M., 1995, ApJ, 445, 80L

Lonsdale, C.J., Barthel, P.D., 1986, ApJ, 303, 617

McCarthy, P.J., 1993, ARA\&A, 31, 639

McCarthy, P.J., van Breugel, W., Kapahi, V.K., 1991, 
 ApJ,  371, 478

Meisenheimer, K., Hippelein, H., 1992, A\&A, 264, 455

Muxlow, T.W.B, Garrington, S.T., 1991, in Hughes, P.A., ed, 
Beams and Jets in Astrophysics,  CUP, Cambridge, p. 51

Nittman, J., Falle, S.A.E.G. and Gaskell, P.H., 
     1982, MNRAS, 201,    833

Norman, M.L., 1993, in Burgarella, D., Livio, M., O'Dea, C., eds, 
STScI Symp. Ser. Vol. 6, Astrophysical Jets, CUP, Cambridge, p. 211

Owen, F.N., O'Dea, C.P., Keel, W.C., 1990, ApJ, 352, 44

O'Donoghue, A.A., Eilek, J.A., Owen, F.N., 1993,  ApJ, 408, 428

O'Donoghue, A.A., Owen, F.N., Eilek, J.A., 1990,  ApJS, 72, 75

Pedelty, J. A., Rudnick, L., McCarthy, P.J., Spinrad, H., 
1989, AJ,    97,      647

Pinckney, J., Burns, J.O., Hill, J.M., 1994,  AJ, 108, 2031

Raga, A.C., Canto, J., 1996, MNRAS, 280, 567

Rector, T.A., Stocke, J.T., Ellingson, E., 1995, AJ, 110, 1492

Rottgering, H., 1993. {\it PhD Thesis}, University of Leiden

Saikia, D.J., Jeyakumar, S., Wiita, P.J., Sanghera, H.,
Spencer, R.E., 1995, MNRAS, 276, 1215

Sakelliou, I., Merryfield, M.R., McHardy, I.M., 1996, MNRAS, 283, 673

Sanders, R.H., Prendergast, K.H., 1974,  ApJ,  188, 489

Scheuer, P.A.G., 1982, in Heeschen, D.S., Wade, C.M., eds, 
Extragalactic Radio Sources, IAU Symp. 97, Reidel, Dordrecht, p. 163

Smith, M., Norman, M.L., Winkler, K-H.A., Smarr, L., 
1985, MNRAS, 214, 67

Sod, Gary A.,  1978,  J.Comp.Phys.,     27,      1      

Steffen, W.S., G\'{o}mez, J.L., Raga, A.C.\ \& Williams, R.J.R, 1997, ApJ, L73

Stocke, J.T., Burns, J.O., Christiansen, W.A., 1985, ApJ, 299, 799

Stockton, A., Ridgway, S., Kellog, M.,  1996, AJ, 112, 902

Sutherland, R.S., Bicknell, G.V., Dopita, M.A., 1993, ApJ, 414, 510

Tadhunter, C.N., Clark, N., Shaw, M.A., Morganti, R., 1994
A\&A, 288, L21

Tadhunter, C.N., 1996, in Carilli, C.L., Harris, D.E.,
Cygnus~A~--~Study~of a Radio Galaxy, CUP, Cambridge, p. 33

van Breugel, W.J.M., Heckman, T.M., George, K., Filippenko, 
A.V.,  1986,  ApJ, 311, 58

van Breugel, W.J.M., 1996, in Ekers, R., Fanti, C., Padrielli, 
L., Extragalactic Radio Sources, IAU Symp. 175, Kluwer, Dordrecht, 577

Williams, A.G., Gull, S.F.,   1985,  Nat,     313,     34

Wilson, Andrew S., 1993, in Burgarella, D., Livio, M., O'Dea, C., eds, 
STScI Symp. Ser. Vol. 6, Astrophysical Jets,  CUP, Cambridge, p. 122

Wilson, M.J., Scheuer, P.A.G.,  1983,  MNRAS,    205,     449

Yates, M.G., Miller, L., Peacock, J.A., 1989,  MNRAS,  240, 129

\newpage

\section*{Figure Captions}

\begin{figure}
\vspace{17.5cm}

\includegraphics{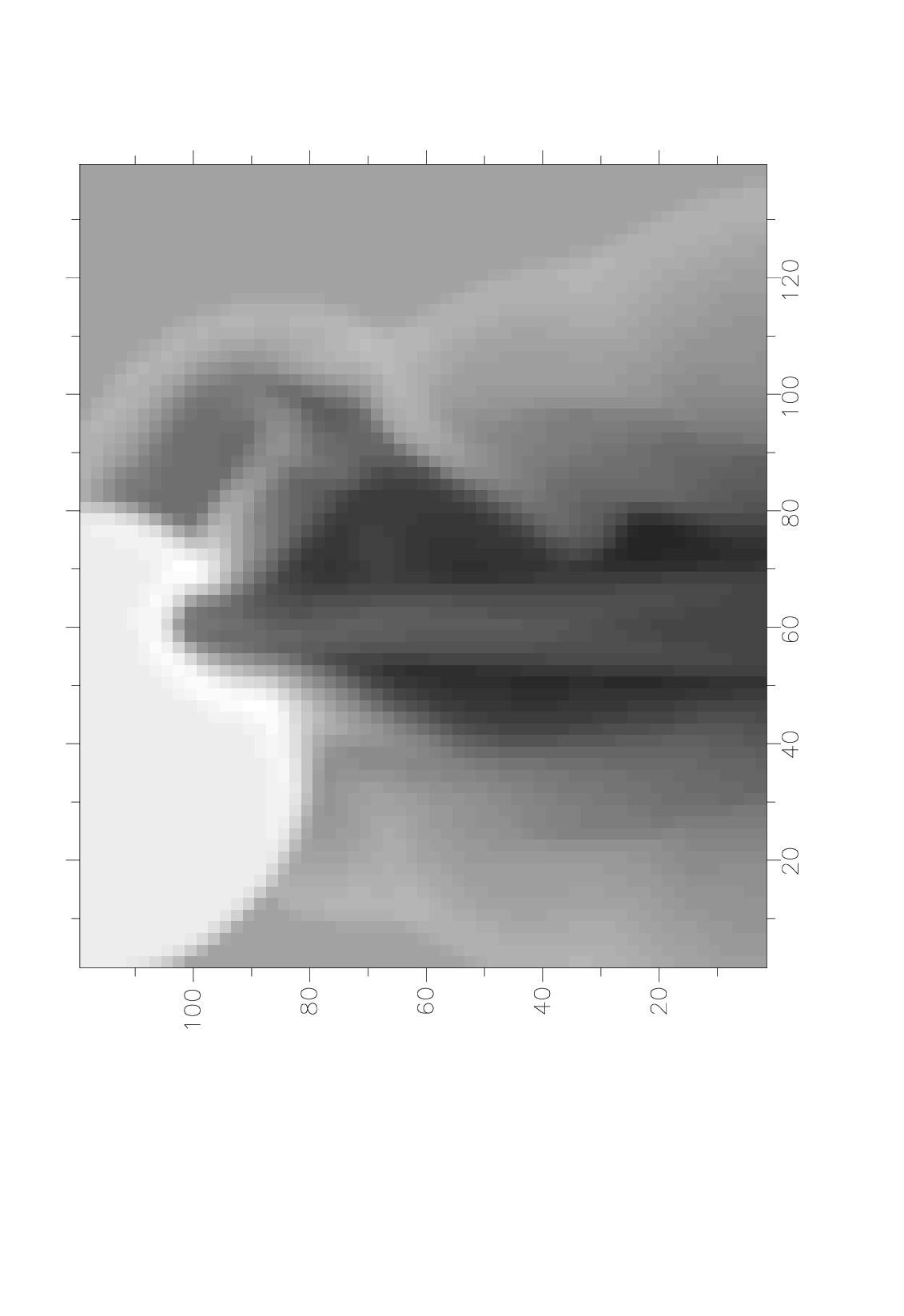}
\includegraphics{jet2050_d-l10.geps}

\vspace{0.5cm}
\caption{}
\label{hirestest}
\end{figure}

\begin{figure}

%\hspace{3mm} 
a) \hspace{8.3cm} b)

\vspace{8cm}

\includegraphics{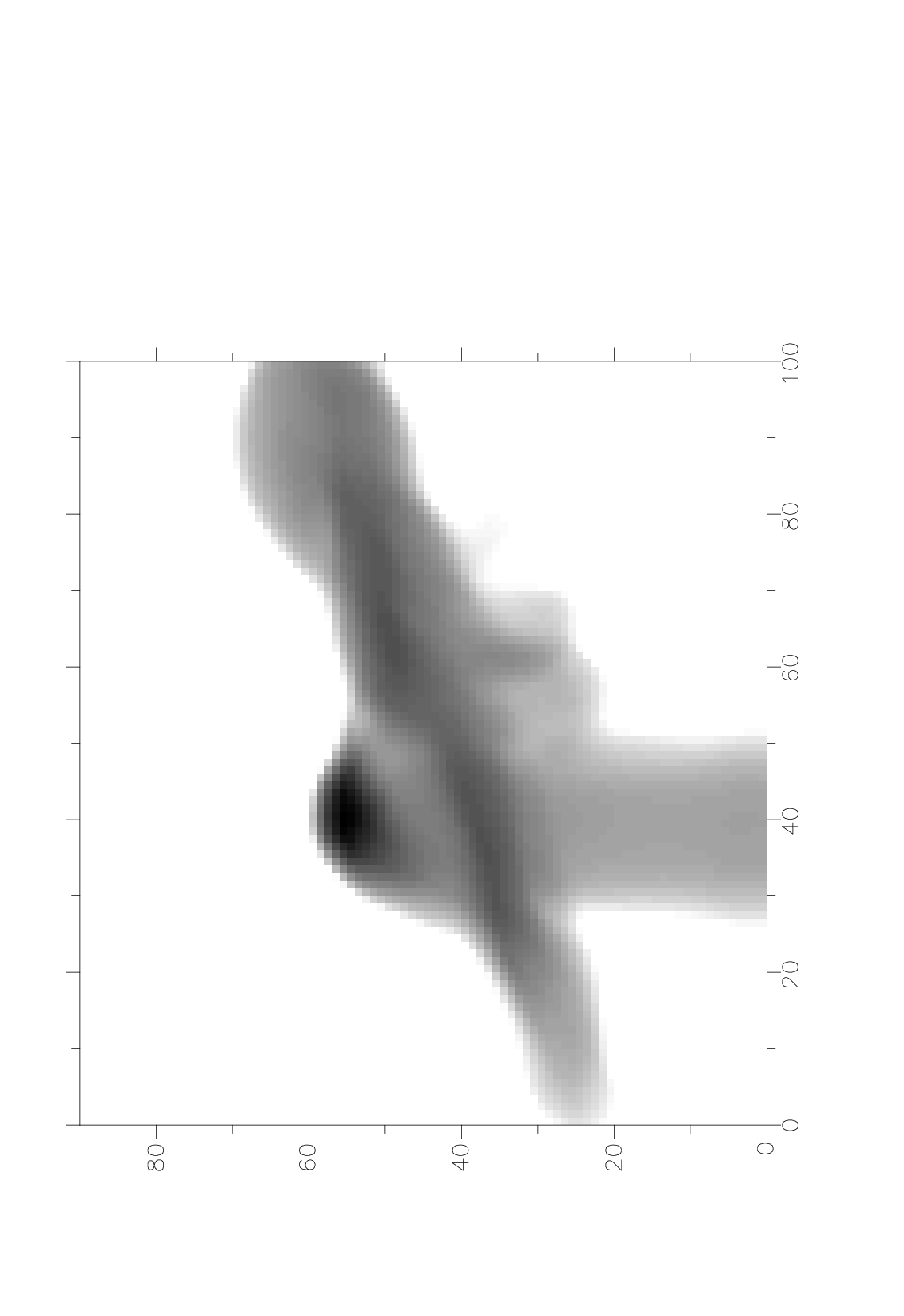}
\includegraphics{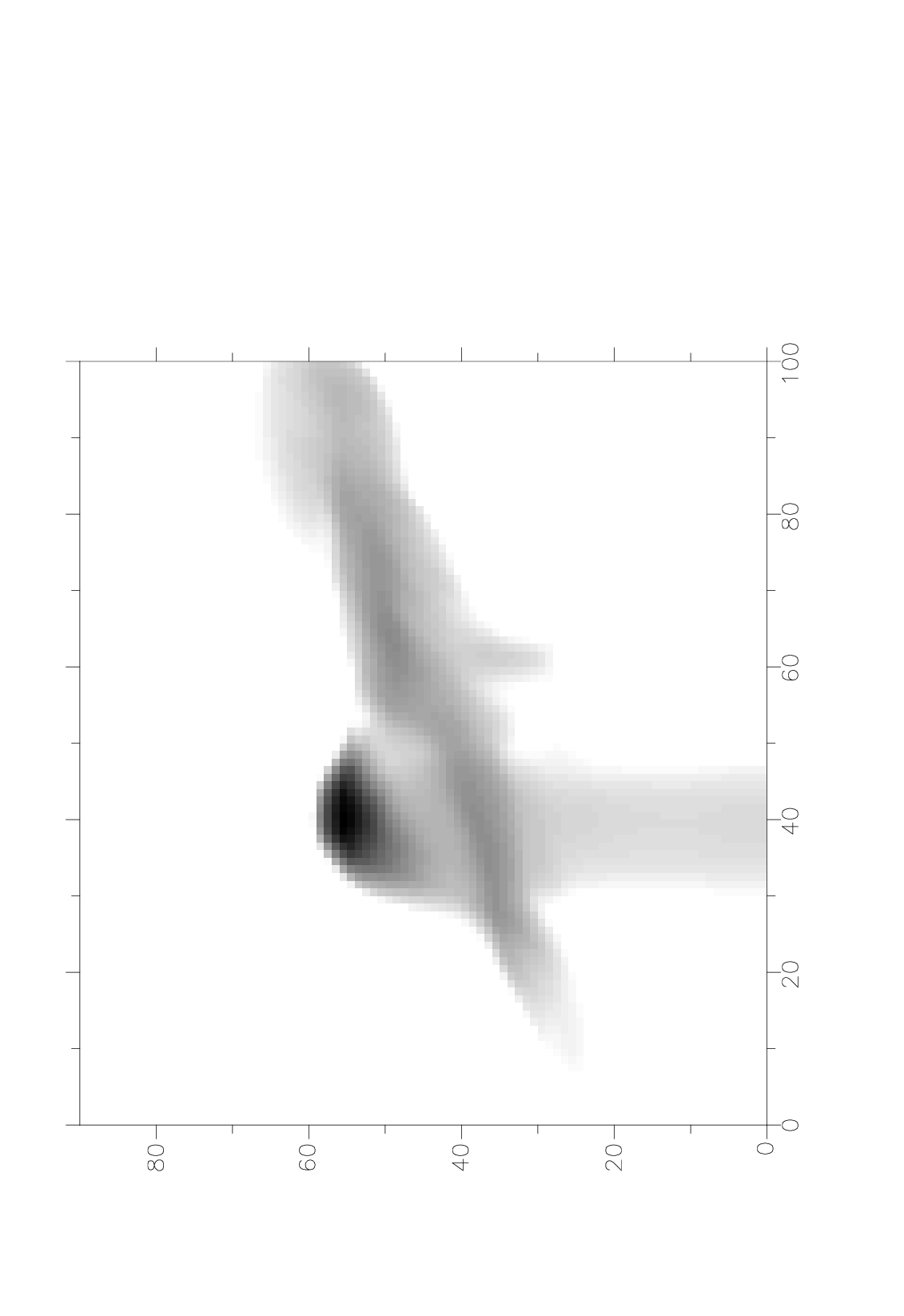}

%\hspace{3mm} 
c) \hspace{8.3cm} d)

\vspace{8.0cm}

\includegraphics{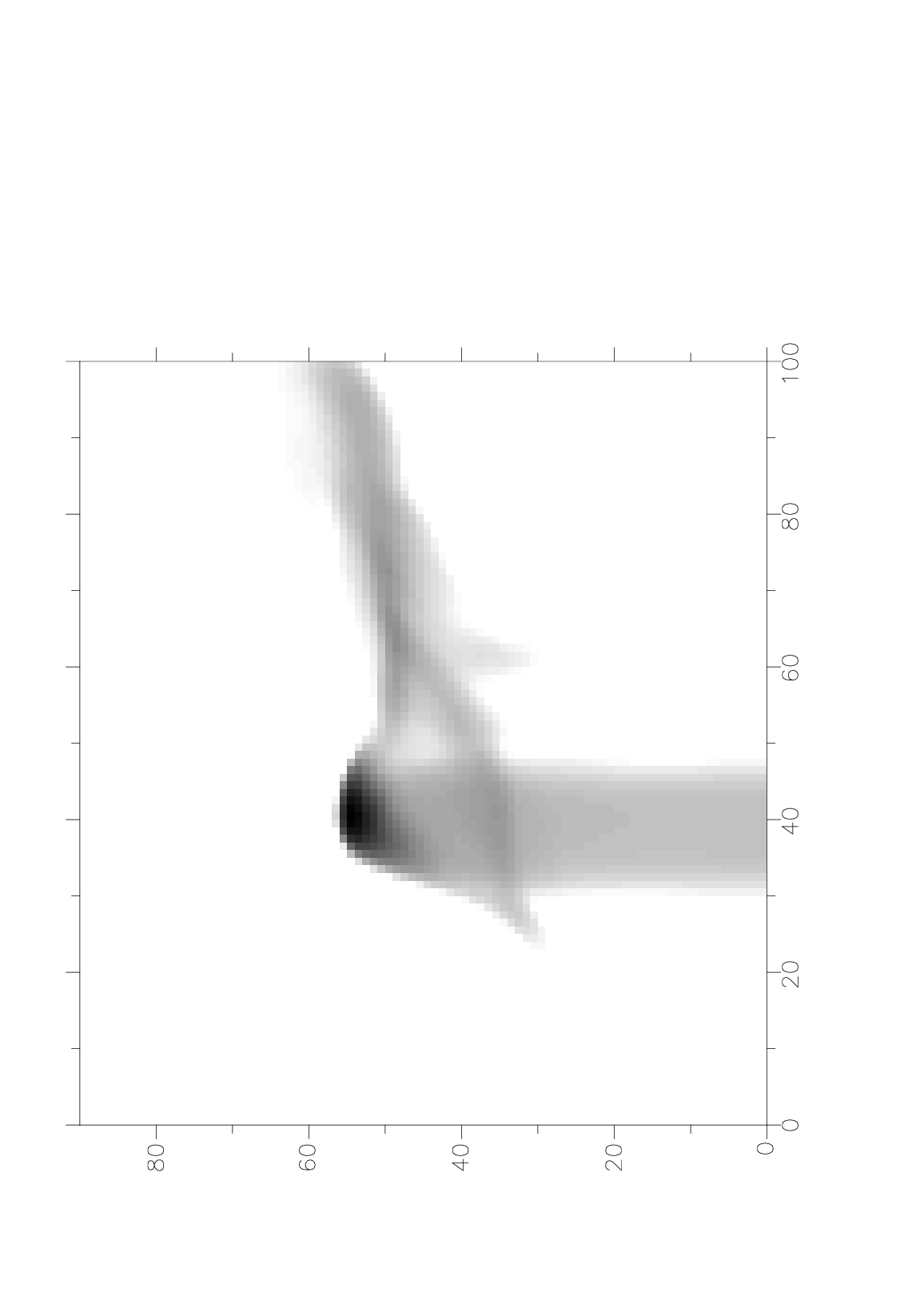} 
\includegraphics{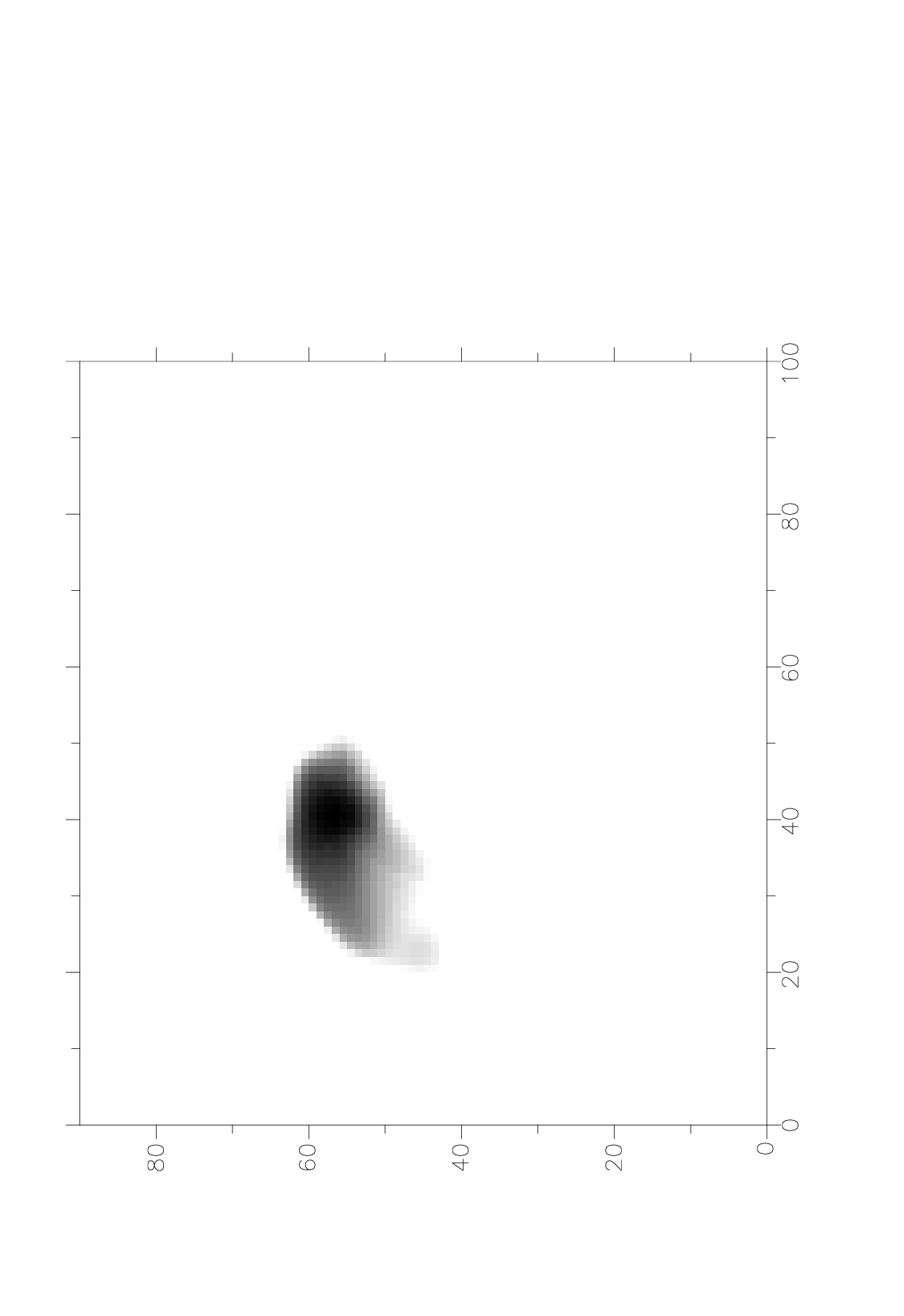} 

\vspace{0.8cm}
\caption{}

\label{rad_comp}

\end{figure}

\begin{figure}

a)

\vspace{10cm}

\includegraphics{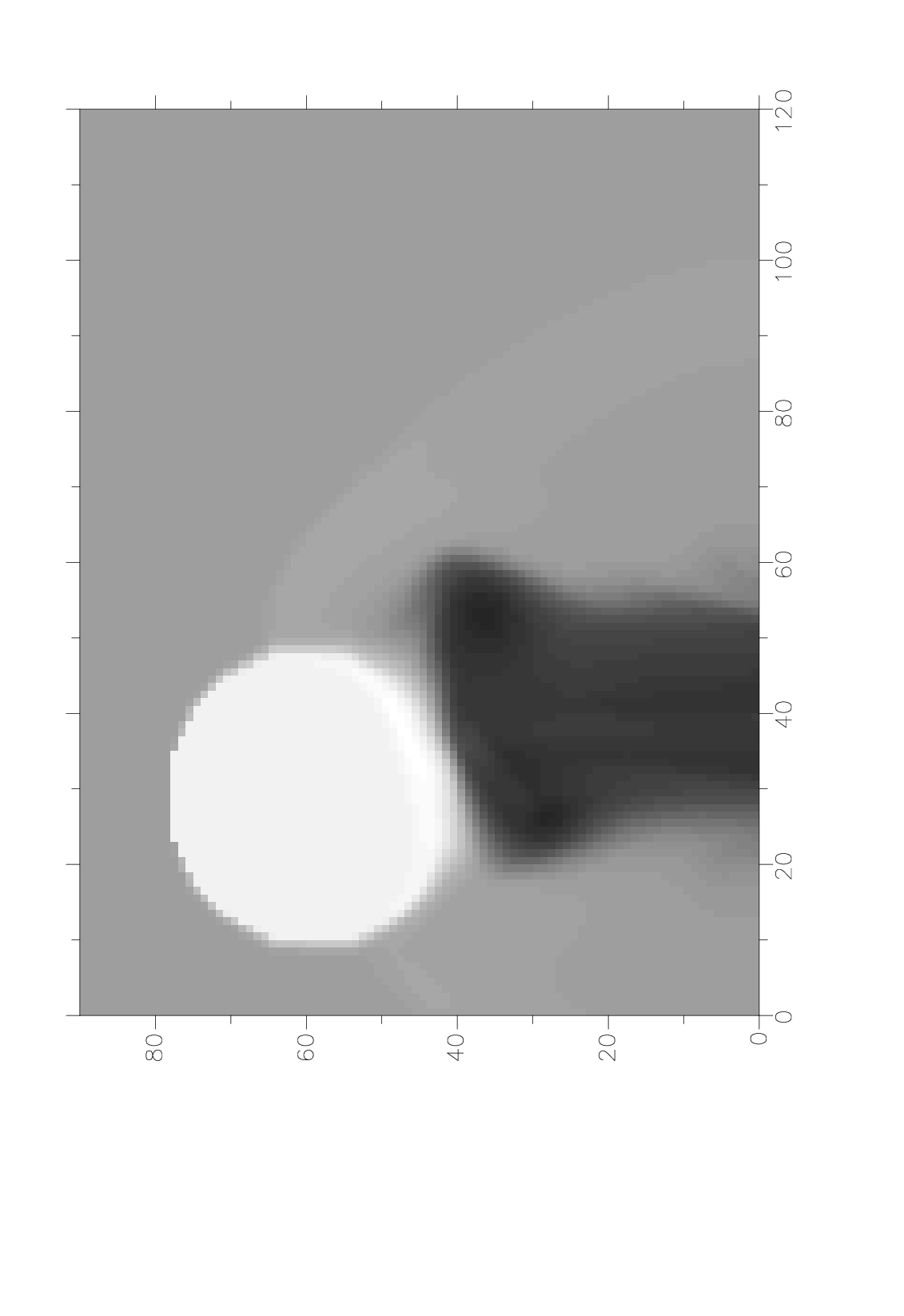}
\includegraphics{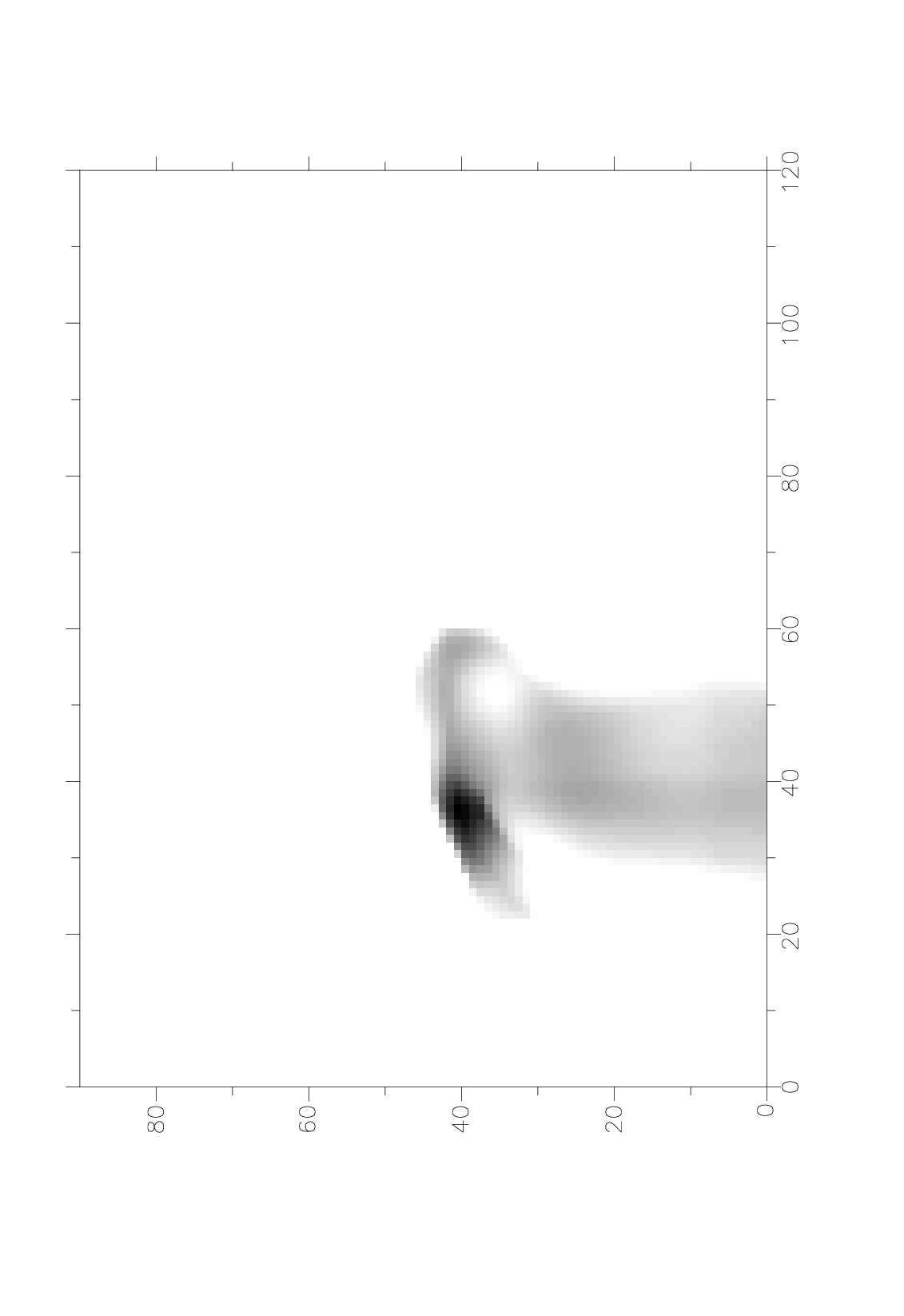}

\vspace{0.5cm}

b)

\vspace{10cm}

\includegraphics{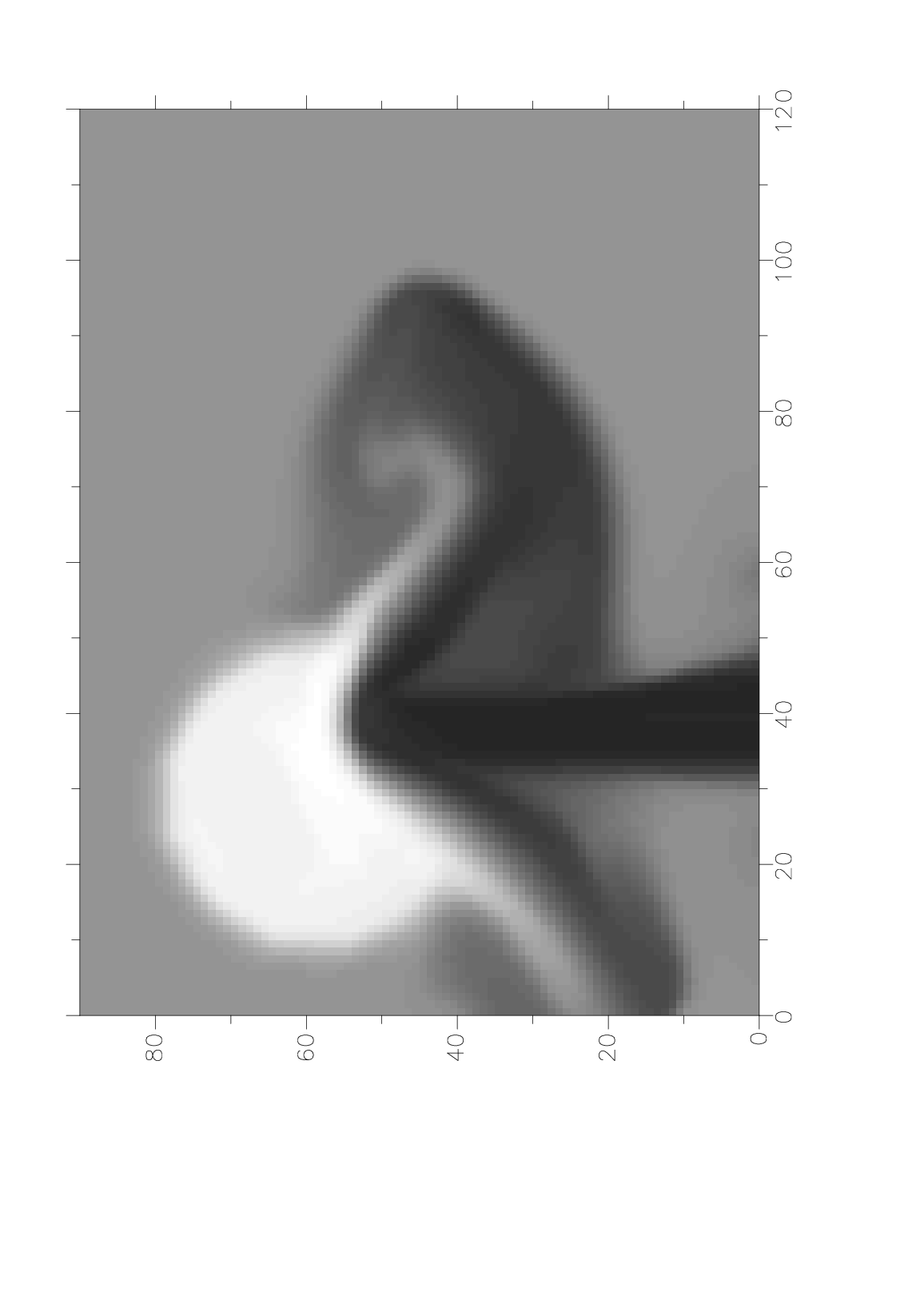}
\includegraphics{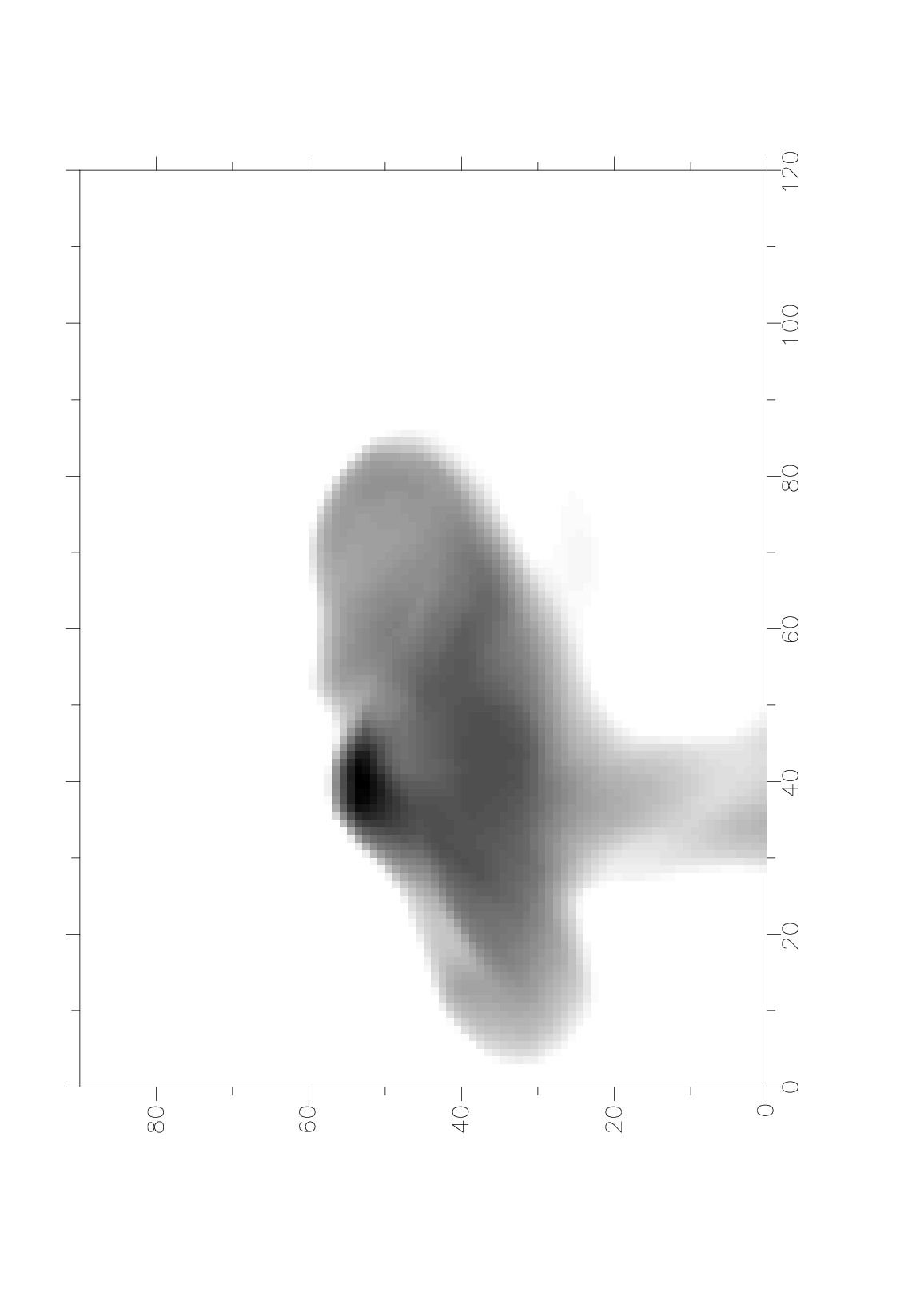}

\caption{}

\label{jet1} 
\end{figure}

\begin{figure}

a)
\vspace{10.5cm}

\includegraphics{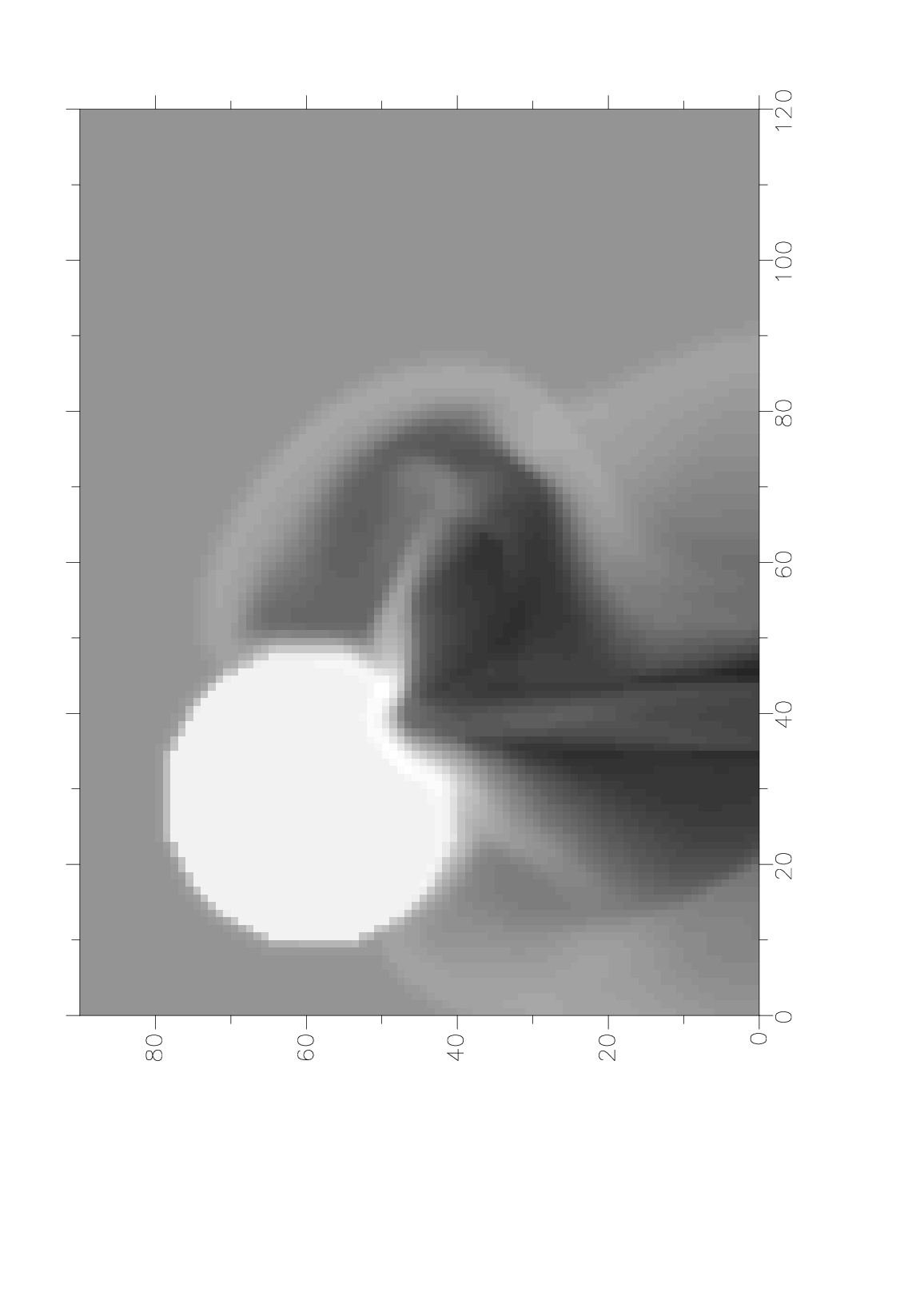}
\includegraphics{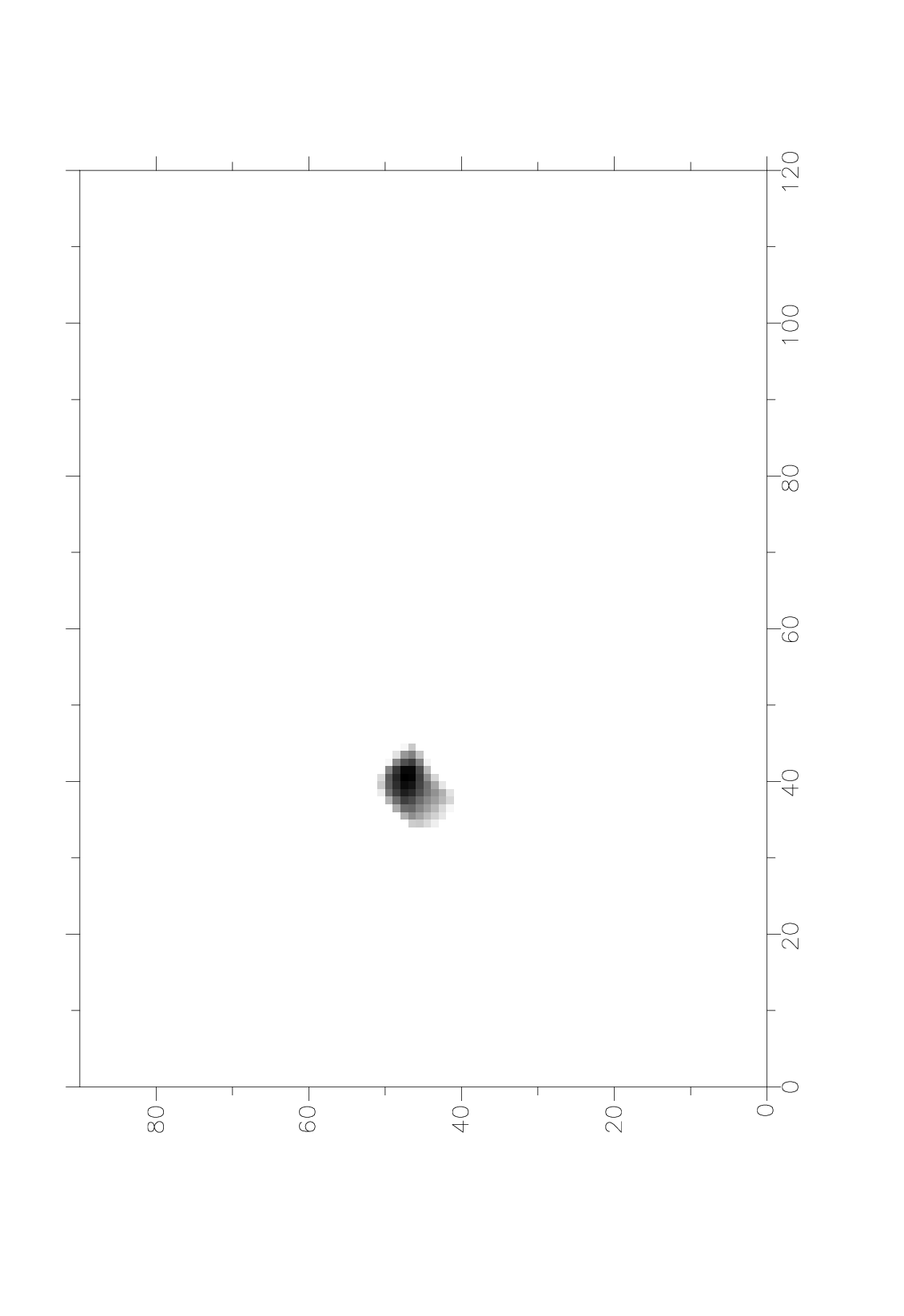}

\vspace{0.5cm}

b)
\vspace{10cm}

\includegraphics{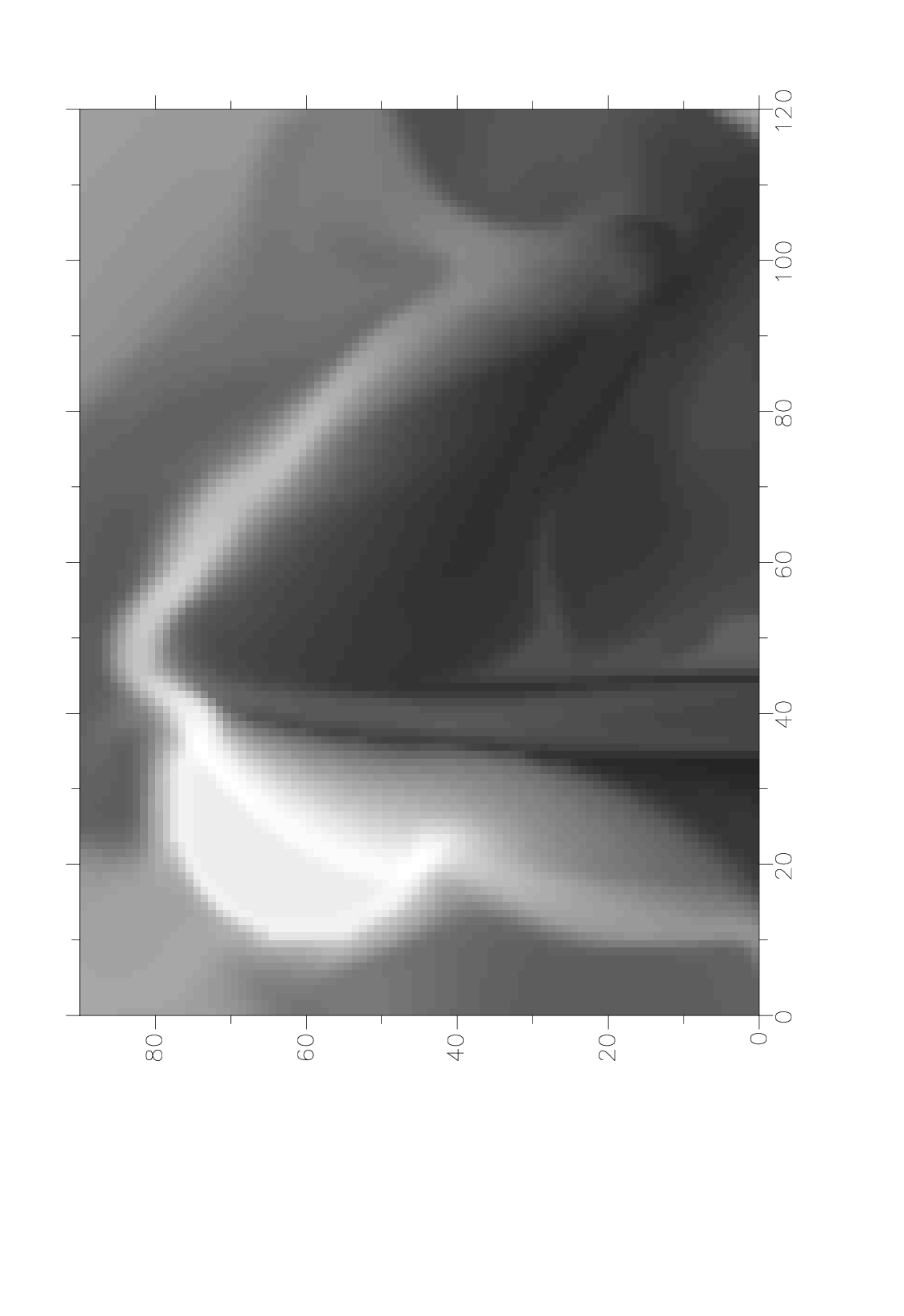}
\includegraphics{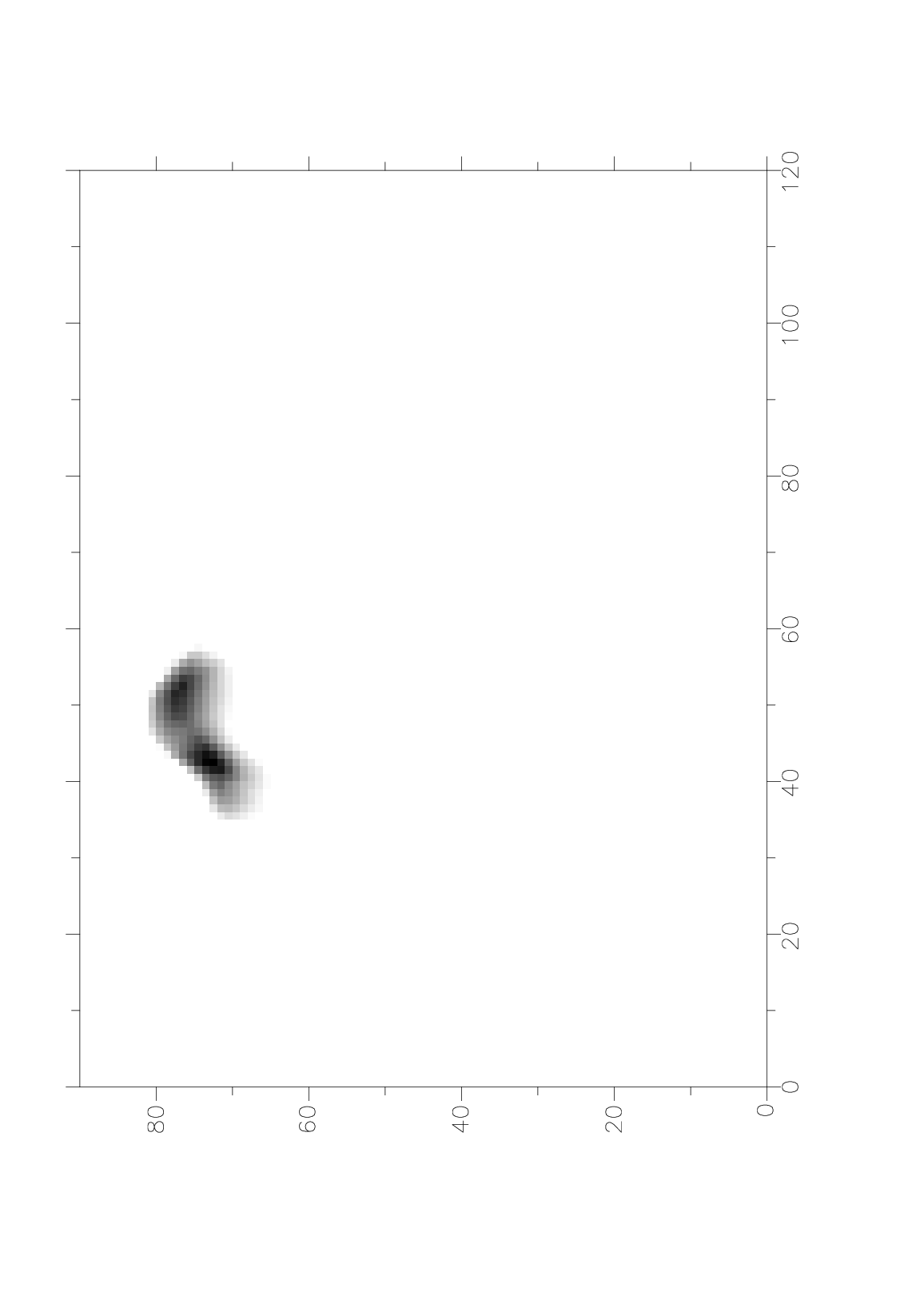}

\caption{}

\label{jet4}
\end{figure}

Figure 1: The geometry of the computational grid. This shows the
symmetry plane containing the jet axis (arrow)  and bisecting the cloud
(hemisphere). It
also shows the viewpoint used for the density slices in figures
\ref{hirestest} to \ref{jet7}, the $z$-axis.

Figure 2: Example of the results of a test 
using a one-dimensional problem involving the impact of a shock on a
density discontinuity. The solid line represents the analytical
solution, the points represent the results of the simulation. The
parameter values are: `jet' (shock) Mach number 2.0, pressure 1.0 and
density 1.0; `cloud' (discontinuity) pressure 1.0 and density 200.

Figure 3: Logarithmic plots (on the same greyscale) of density in the
symmetry plane for the same jet parameters at the same time run at two
different resolutions. The upper panel was produced from a simulation
run at twice the resolution of the second, then smoothed with a 3-D
gaussian and binned down to the same pixel size.

Figure 4: Alternative plots of synchrotron surface brightness distribution, 
using a logarithmic grey scale. Panels {\em a)} and {\em b)} show the effect
of changing $\alpha$ on our prescription. Panel {\em a)}~shows the effect
of $\alpha$ = 0.5, which is the value we use in plots shown elsewhere
in this paper. Panel {\em b)}~shows the effect of $\alpha$ = 1.0: jet and
tail are fainter, so that the wider diffuse emission at the edges of
the tail is not seen. Panel  {\em c)}~shows pressure squared only where
there is jet material. The jet and tail are as bright as in the
previous plots, but the diffuse emission around them is not seen due
to the negligible magnetic fields in these outer regions, producing a
narrower tail. Panel  {\em d)}~shows the square of total pressure: this
is the effect of allowing the same magnetic field in the ambient
medium and the jet.

Figure \ref{jet1}: Logarithmic greyscale plots of total gas 
density (in the symmetry plane)
and integrated radio emission form simulation 1 ($M=2,\eta_j=0.01,\eta_c=50.0$). 
In the density plots white represents the highest value. For clarity in the radio
plots black represents the peak brightness. The plots are shown at two epochs from
the simulation: {\it a)} 56 million years after the jet enters the grid, and {\it
b)} 112 million years.

Figure \ref{jet4}: Logarithmic greyscale plots of total gas 
density (in the symmetry plane)
and integrated radio emission for simulation 4 ($M=10, \eta=0.01, \eta_c=200.0$).
In the density plots white represents the highest value. For clarity in the radio
plots black represents the peak brightness.
The plots are shown at two epochs from the simulation: {\it a)} 16 million years
after the jet enters the grid, and {\it b)} 56 million years.

Figure \ref{jet7}: Logarithmic greyscale plots of total gas 
density (in the symmetry plane)
and integrated radio emission for simulation 6 ($M=2, \eta_j=0.2, \eta_c=200.0$). 
In the density plots white represents the highest value. For clarity in the radio
plots black represents the peak brightness.  The plots are shown at two epochs
from the simulation: {\it a)} 56 million years after the jet enters the grid, and
{\it b)} 112 million years.

Figure \ref{jet7_1}: Three dimensional velocity vector plots of a 
Mach 2 jet 
with density contrast $\eta=0.2$, seen from 
two orientations. The jet 
emerges from the plane of the upper panel. Vectors are plotted if they
have a speed greater than 20\% of the speed of the incoming jet.

Figure \ref{rad}: Radio emission seen from two 
orientations:{\em (a)}~perpendicular
to the symmetry plane (as in previous radio plots), 
and {\em (b)}~rotated 
 so that the jet axis is about 30$^{\circ}$ from the line of sight.

Figure \ref{em}: Square of density integrated along the line of sight, 
indicating 
the expected site of line emission, overlaid with synthesized radio
contours corresponding to \ref{rad}~{\em a)}.

Figure \ref{multi}: A jet propagating through a collection of clouds. 
This is a fast, light jet 
($M$=10, $\eta$=0.01). All clouds have density contrast of 50. They are
distributed at random positions in the grid, with a fixed volume filling factor of
0.2 and a power law distribution of radius up to a fixed maximum size (1.4 times
the radius of the jet). There is no plane of symmetry in this problem, so we had
to calculate the whole grid, which in this case was 90 $\times$ 90 $\times$ 90.
This represents a volume of 729000 kpc$^3$. The plots show the results about four
million years after the jet entered the grid.  {\em a)}~A constant density surface
of the jet inside a diffuse rendering of cloud density. {\em b)}~A
synthetic radio map. The jet has been deflected at least twice, where it has
encountered clouds, but clearly remains collimated.

\newpage

\onecolumn

\begin{table}

\begin{tabular}{lllll}
Simulation & Jet density & Cloud density & Jet mach & Jet \\
number     & contrast    & contrast      & number    &  speed         \\
\hline
1 & 0.01  & 50  & 2  & 0.07c \\
2 & 0.01  & 200 & 2  & 0.07c \\
3 & 0.01  & 50  & 10 & 0.36c \\
4 & 0.01  & 200 & 10 & 0.36c \\
5 & 0.2   & 50  & 2  & 0.02c \\
6 & 0.2   & 200 & 2  & 0.02c \\
7 & 0.2   & 50  & 10 & 0.08c \\
8 & 0.2   & 200 & 10 & 0.08c \\
9 & 0.01  & --  & 2  & 0.07c \\
10 & 0.01 & --  & 2  & 0.02c \\
11 & 0.2  & --  & 10 & 0.36c \\
12 & 0.2  & --  & 10 & 0.08c \\
\end{tabular}
\caption{The values of the parameters for twelve different simulations. The
jet density contrast is the ratio of the jet density to the ambient
 density; the cloud density contrast is the ratio of the cloud density
 to the ambient density; the jet mach number is the ratio of the jet
 speed to the sound speed within the jet. There is no cloud in
 simulations 9 -- 12.}

\label{pstbl}
\end{table}

\newpage

\newpage

\newpage

\begin{figure}

a)
\vspace{10cm}

\includegraphics{jet7280_d-l10.ps}

\includegraphics{jet7280_rad2-l10.ps}

\vspace{0.5cm}

b)
\vspace{10cm}

\includegraphics{jet7760_d-l10.eps}
\includegraphics{jet7760_rad2-l10.eps}

\caption{}

\label{jet7} 
\end{figure}

\newpage

\begin{figure}

\vspace{23cm}

\includegraphics{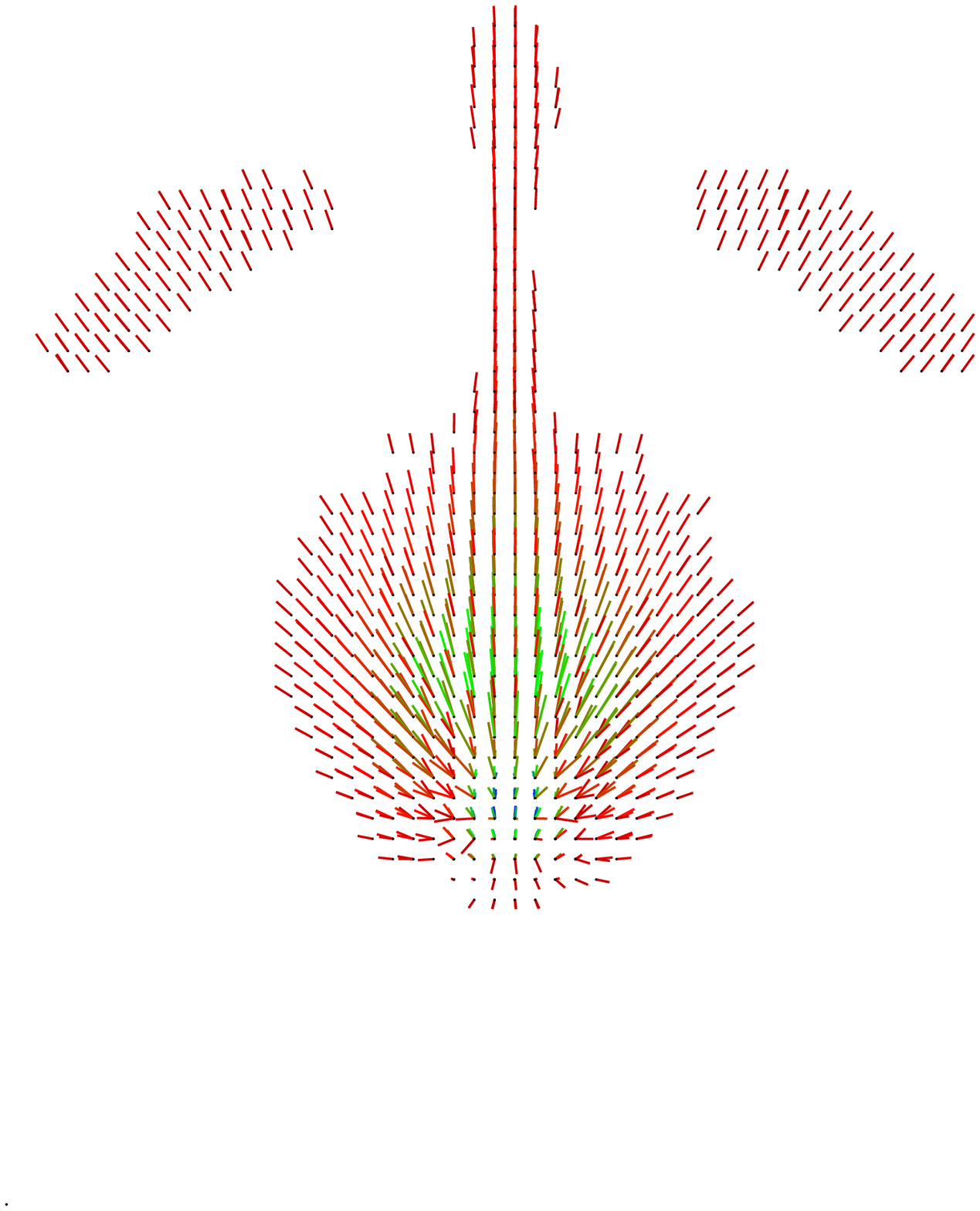} 
\includegraphics{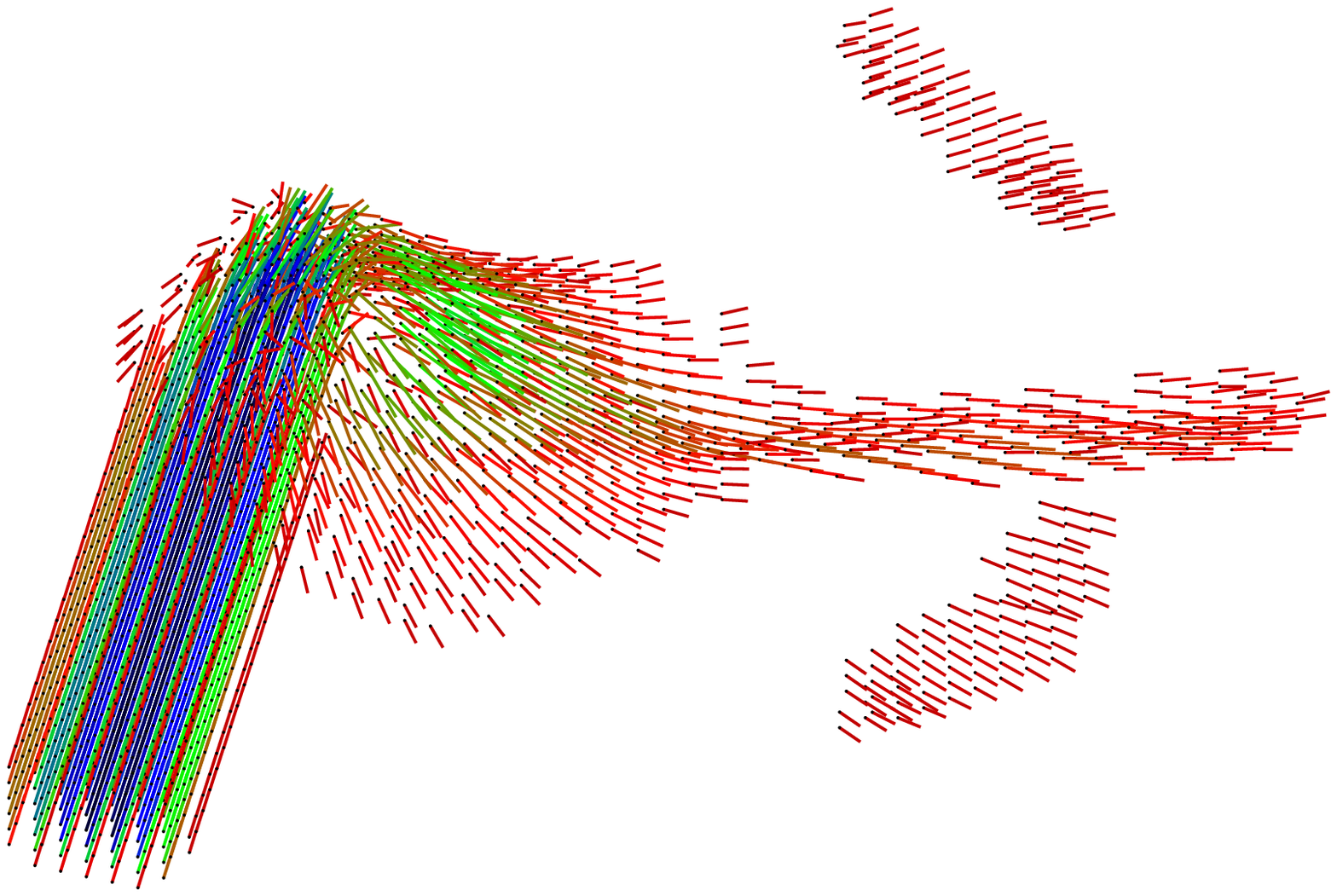} 

\caption{}

\label{jet7_1}

\end{figure}
\newpage

\begin{figure}

\vspace{1cm}
a)
\vspace{10cm}

\includegraphics{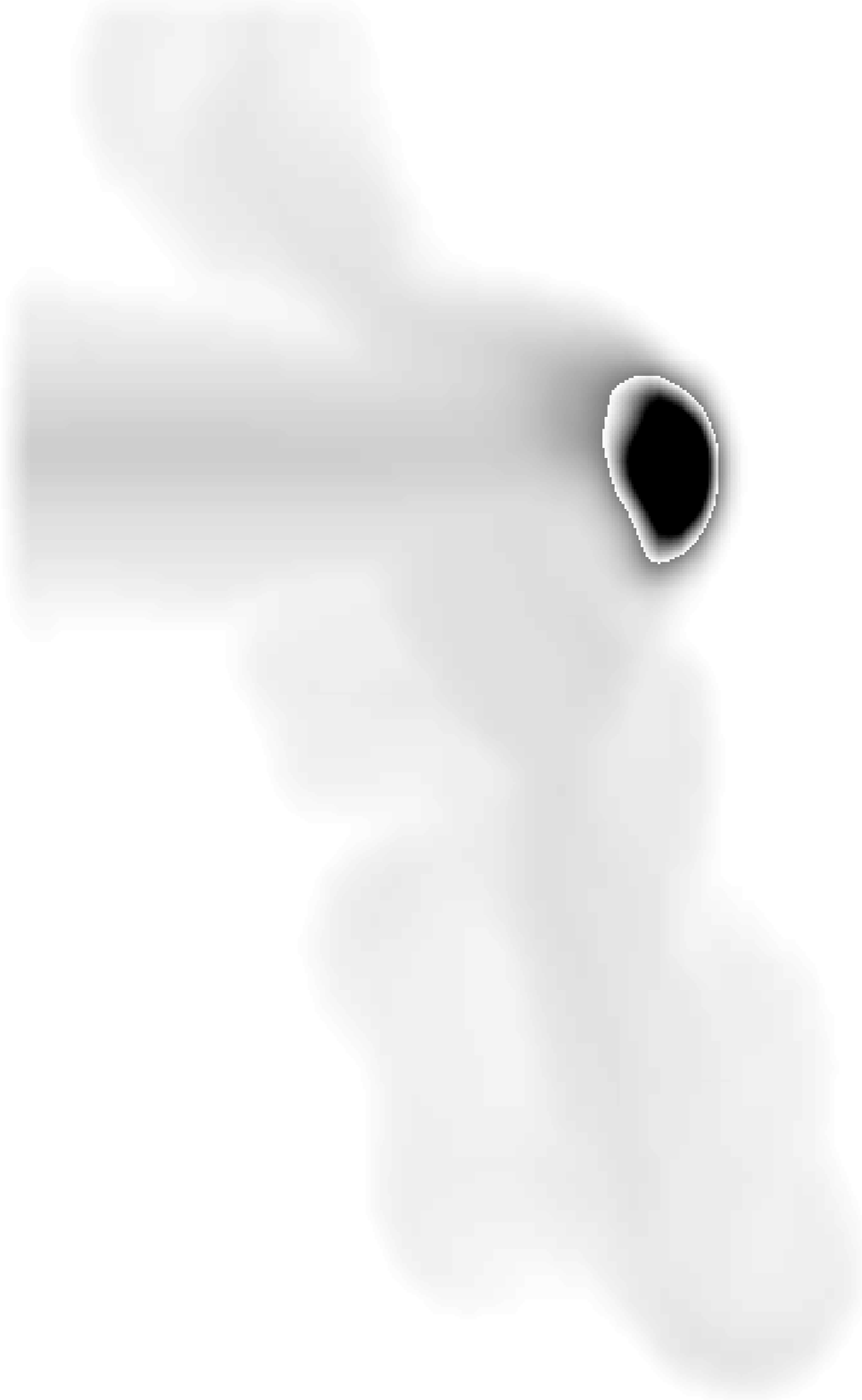}

b)
\vspace{11cm}

\includegraphics{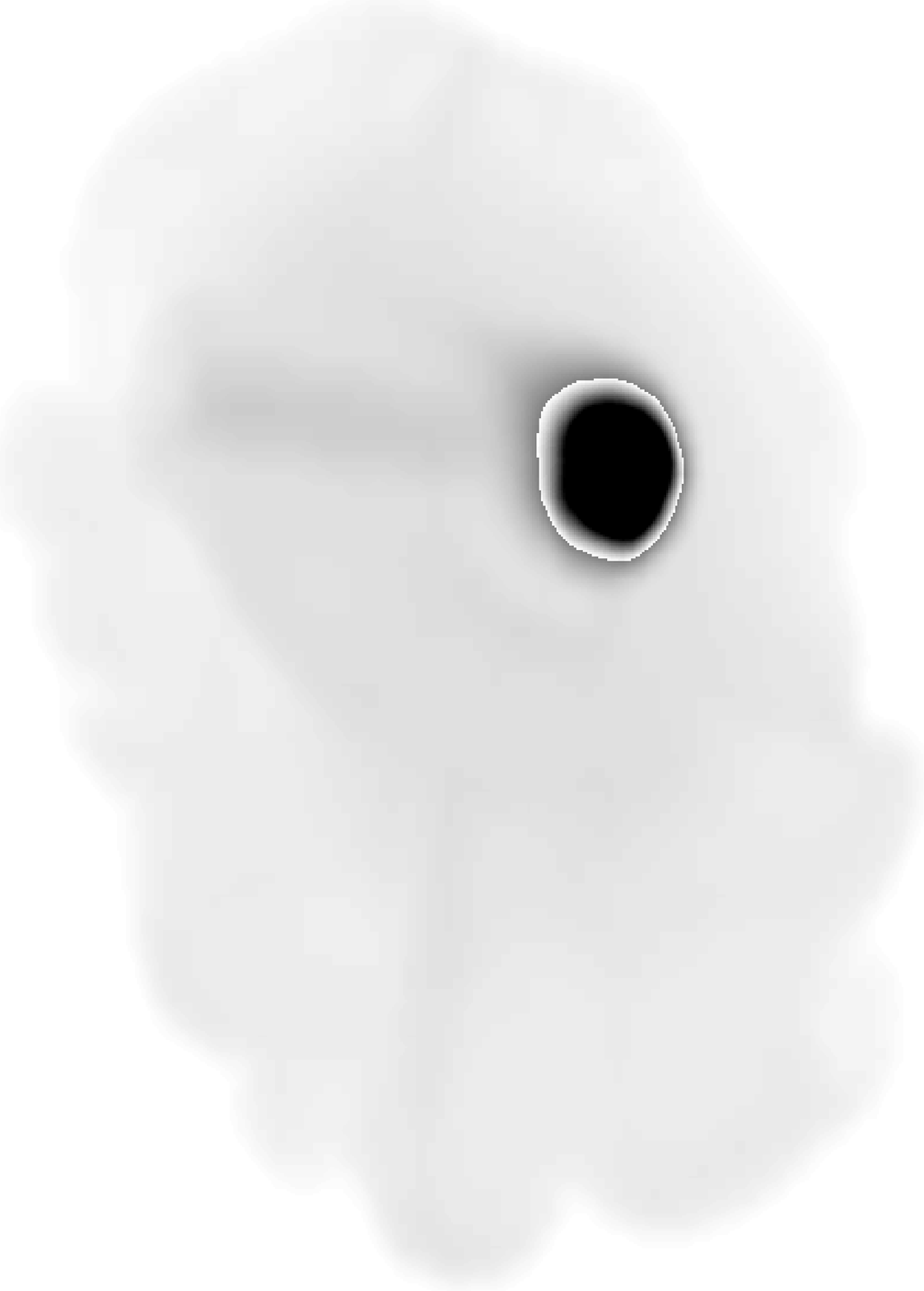}

\caption{}

\label{rad}

\end{figure}
\newpage

\begin{figure}

\vspace{12cm}

\includegraphics{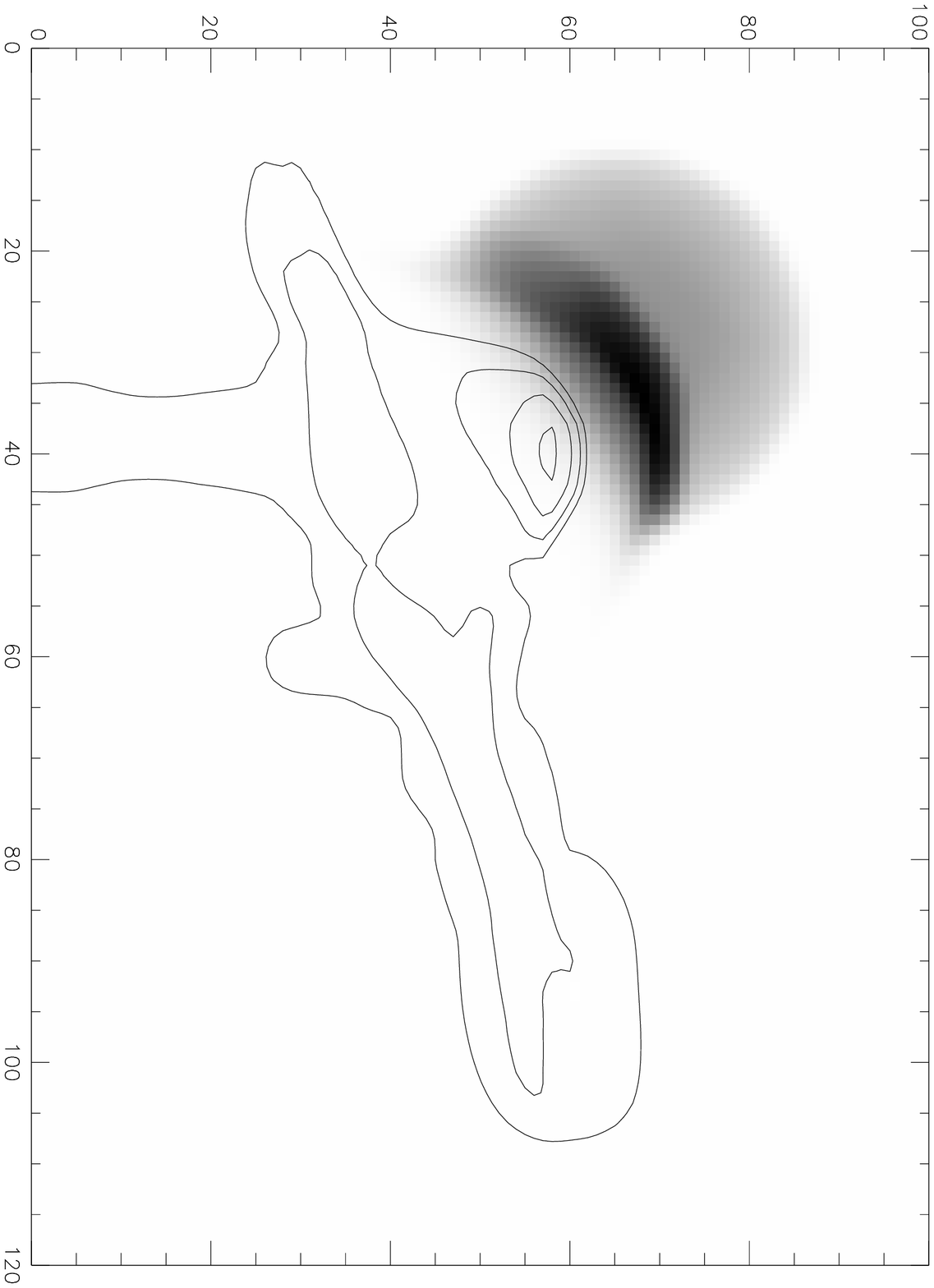}

\caption{}

\label{em}
\end{figure}

\begin{figure}

 a)

\vspace{11cm} b)

\vspace{10cm}

\includegraphics{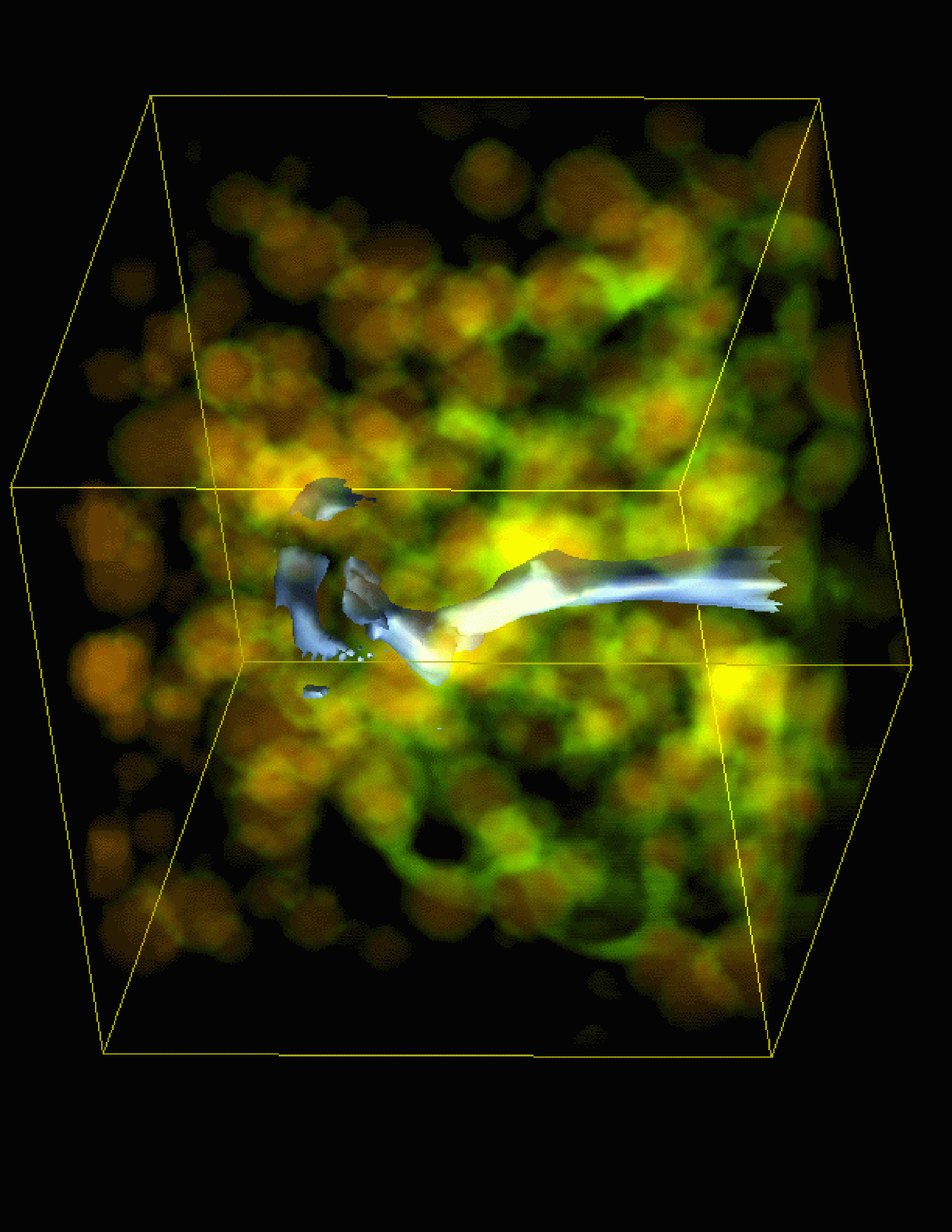} 
\includegraphics{jetrnd080_rad5.eps}

\caption{}

\label{multi}

\end{figure}

\end{document}